\documentclass[prb,reprint,superscriptaddress]{revtex4-2}
\usepackage[utf8]{inputenc}
\usepackage{amsmath}
\usepackage[colorlinks=true,allcolors={blue}]{hyperref}
\usepackage{graphicx}
\usepackage{hhline}% http://ctan.org/pkg/hhline
\usepackage{multirow}% http://ctan.org/pkg/multirow
\usepackage{tabularx} 
\usepackage{array}
\usepackage{makecell}
\usepackage{braket}
\usepackage[normalem]{ulem}
\usepackage{ulem}
\usepackage{siunitx,booktabs}
\usepackage{comment}
\usepackage{color}
\usepackage{amsmath}
\makeatletter
\newcommand{\ph}[1]{\phantom{#1}}
\newcommand{\up}{\uparrow}
\newcommand{\dn}{\downarrow}
\newcommand{\mat}[1]{{\mathpalette\mat@{#1}}}
\newcommand{\mat@}[2]{%
  \begingroup
  \sbox\z@{$\m@th#1\underline{#2}$}%
  \dimen@=\dp\z@ \advance\dimen@ -2\mat@dimen{#1}%
  \dp\z@=\dimen@
  \sbox\z@{$\m@th\underline{\box\z@}$}%
  \box\z@
  \endgroup
}
\newcommand\mat@dimen[1]{%
  \fontdimen8
  \ifx#1\displaystyle\textfont\else
  \ifx#1\textstyle\textfont\else
  \ifx#1\scriptstyle\scriptfont\else
  \scriptscriptfont\fi\fi\fi 3
}
\begin{document}
%\linenumbers

\author{Sergii Grytsiuk}
\affiliation{Institute for Molecules and Materials, Radboud University, Heijendaalseweg 135, 6525AJ Nijmegen, The Netherlands}

\author{Mikhail I. Katsnelson}
\affiliation{Institute for Molecules and Materials, Radboud University, Heijendaalseweg 135, 6525AJ Nijmegen, The Netherlands}

\author{Erik G.C.P. van Loon}
\affiliation{Division of Mathematical Physics, Physics Department, Lund University, Sweden}

\author{Malte R\"osner}
\affiliation{Institute for Molecules and Materials, Radboud University, Heijendaalseweg 135, 6525AJ Nijmegen, The Netherlands}

\title{Nb$_3$Cl$_8$: A Prototypical Layered Mott-Hubbard Insulator}

\begin{abstract}
    The Hubbard model provides an idealized description of electronic correlations in solids. Despite its simplicity, the model features a competition between several different phases that have made it one of the most studied systems in theoretical physics. Real materials usually deviate from the ideal of the Hubbard model in several ways, but the monolayer of Nb$_3$Cl$_8$ has recently appeared as a potentially optimal candidate for the realization of such a single-orbital Hubbard model. 
    Here we show how this single orbital Hubbard model can be indeed constructed within a ``molecular'' rather than atomic basis set using \textit{ab initio} constrained random phase approximation calculations. This way, we provide the essential ingredients to connect experimental reality with \textit{ab initio} material descriptions and correlated electron theory, which clarifies that monolayer Nb$_3$Cl$_8$ is a Mott insulator with a gap of about $1$ to $1.2$eV depending on its dielectric environment. By comparing with an atomistic three-orbital model, we show that the single molecular orbital description is indeed adequate. We furthermore comment on the expected electronic and magnetic structure of the compound and show that the Mott insulating state survives in the low-temperature and bulk phases of the material.
\end{abstract}

\maketitle

\section{Introduction}

  The first theoretical studies of correlation effects in solids date back as early as 1934 with studies on the so-called polar model~\cite{polar1934}. In this model, the crystal was represented as a periodic array of single-orbital sites such that each site can be empty, double occupied, or single occupied with a spin up or down electron, and various types of hopping and interaction processes were taken into account~\cite{VKI1979,VKII1979}. 
  Further simplification resulted in the appearance of the Hubbard model~\cite{Gutzwiller1963,Kanamori1963,Hubbard1963,Hubbard1964,HubbardIII} in the 1960s, in which only basic processes of single-electron hopping and on-site Coulomb repulsion are taken into account. Although the Hamiltonian of this model is simple, the physics that emerges from the competition between interaction and kinetic hopping, or between localization and delocalization, is extremely rich and complicated. As a function of temperature, lattice structure, and number of electrons, the phase diagram of the model is believed to contain metallic and insulating phases, magnetic ordering, and more exotic phases such as unconventional superconductivity and charge-density waves~\cite{Qin22}. In general, the model is too difficult to be solved exactly in the thermodynamic limit (except in one~\cite{Lieb1968} and infinite dimensions~\cite{Metzner89,Georges92}), but many efficient computational strategies have been devised, and some regions of the phase space are well-understood nowadays~\cite{LeBlanc15,Arovas22,Qin22}.

  Given its theoretical importance and computational difficulty, experimental realizations of the Hubbard model have long been sought since then nature itself would solve the model for us. Some unconventional superconductors, including cuprates~\cite{OKA,Pavarini} and nickelates~\cite{Lechermann20,Kitatani20}, as well as Na$_x$CoO$_2$~\cite{Lechermann2015} have been described using the single-orbital Hubbard model. However, in these cases, most simplified single-orbital models are controversial, and there are indications that multi-orbitals models are needed~\cite{Qin22}. Single-orbital models seem to be, nevertheless, adequate to describe transition metal dichalcogenides undergoing charge-density-wave transitions, with one low-energy orbital per supercell~\cite{Perfetti03}. Artificial lattices, in the form of ad-atoms on surfaces~\cite{Hansmann13,Hansmann13b} or atoms trapped in optical lattices~\cite{Esslinger10} are other prominent examples. However, an actual solid that features the ideal realization of Hubbard model physics in the unit cell is of course even more desired. There are several difficulties associated with finding such material. First of all, most materials have several bands relatively close to the Fermi level, which casts doubt on a single-orbital low-energy description. Secondly, background screening in the solid should be so efficient that the Coulomb interactions between electrons on different sites are effectively zero, which is unrealistic for many two-dimensional materials~\cite{Schuler13}. Thirdly, the Coulomb interaction between electrons on the same site should be strong enough to find the desired strong correlation effects. 

  It has been proposed that monolayer Nb$_3$Cl$_8$~\cite{wang_quantum_2023} is a suitable candidate for a true Hubbard material~\cite{Gao2022,Zhang23}. At the DFT level, it has a single well-isolated band crossing the Fermi level. At the same time,  experimental angle-resolved photo emission observations~\cite{sun_observation_2022,Gao2022} and transport measurements~\cite{Yoon2020} show a gap in bulk structures, which could be an indication of strong correlation effects. However, until now, no first-principles calculations of the strength of the local and non-local Coulomb matrix elements have been performed. Here, we use the constrained Random Phase Approximation (cRPA)~\cite{Aryasetiawan04} to calculate the relevant partially-screened Coulomb matrix elements and show that the local Hubbard interaction is indeed sufficient to open a Mott gap in the monolayer. Furthermore, by comparing models with three and one orbitals per unit cell, we demonstrate that the single-orbital model is indeed applicable. We provide all parameters, including interactions and hopping matrix elements, as necessary for many-body calculations of the material's electronic properties and investigate the derived models on the level of the Hubbard-I approximations. Furthermore, we show that the bulk structure can also be described by a slightly more involved two-orbital Hubbard model and investigate it in various experimentally observed high- and low-temperature structures.

    \begin{figure}
        \centering
        \includegraphics[width=0.45\textwidth]{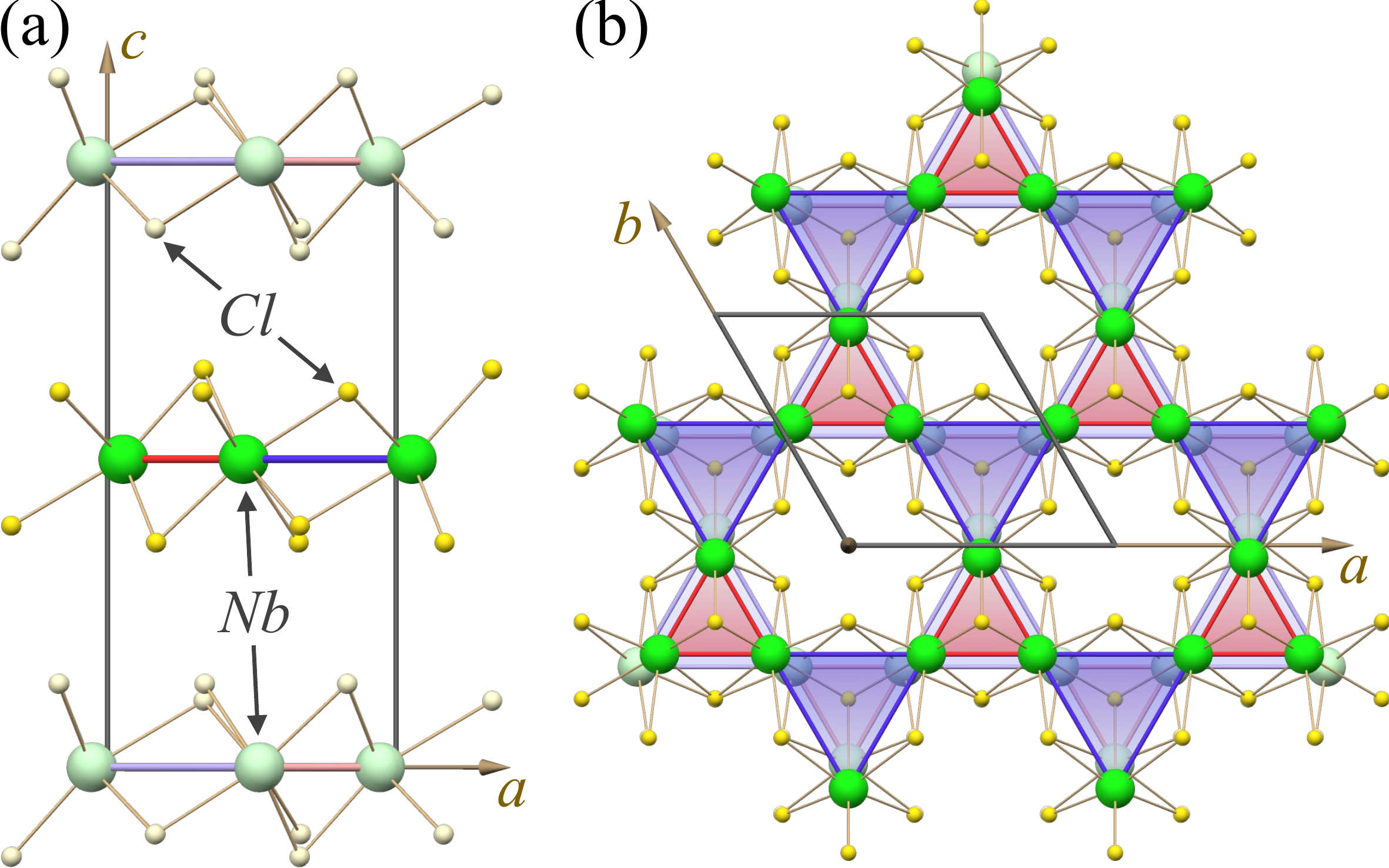}
        \caption{Lattice structure of Nb$_3$Cl$_8$. 
        (a) shows a side view, and 
        (b) shows a top view. 
        Red (blue) triangles indicate the first (second) nearest Nb neighbors, while 
        brighter and dimmer spheres represent atoms (of Nb and Cl) in opposite layers.}
        \label{fig:latticestructure}
    \end{figure}

    \begin{figure*}
        \centering
        \includegraphics[width=0.95\textwidth]{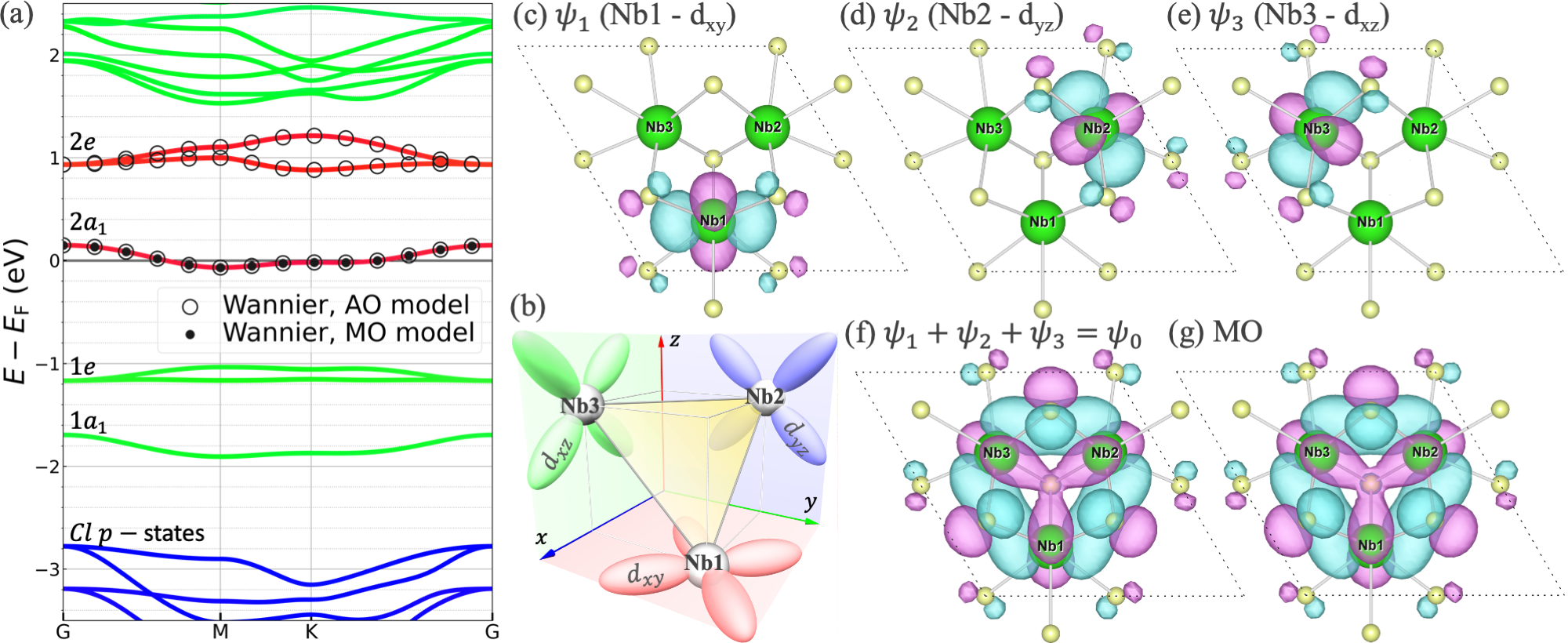}
        \caption{Electronic structure and Wannier orbitals of the atomic and molecular orbital models.
        (a) Shows DFT and Wannier-interpolated electronic structures without spin polarization. Bands in red, green, and blue colors stand for 
        Nb-$t_{2g}^1$, Nb-$(t_{2g}^2 + e_g)$, and Cl $p$ states, respectively. $t_{2g}^1$ orbitals are schematically illustrated in (b) for each Nb atom (in rotated coordinates), while $t_{2g}^2$ represents all other $t_{2g}$ orbitals. 
        Opened  and filled markers stand for the atomic (AO) and 
        molecular (MO) orbital models, respectively.
        (c-e) Show Wannier atomic orbitals for each Nb atom,  (f) their sum, and (g) the Wannier molecular orbital.}
        \label{fig:bandstructure}
    \end{figure*}

\section{Results}

  The high-temperature lattice structure of monolayer Nb$_3$Cl$_8$ is shown in Fig.~\ref{fig:latticestructure}. The Nb atoms form a distorted kagome lattice, with small (red) and big (blue) triangles. The distortion is 16.6\%, i.e., the sides of the small triangle are 16.6\% shorter and those of the big triangle 16.6\% longer than for an undistorted kagome lattice, in which both would be 
  $a/2= 3.36$ \AA. 
  Every unit cell contains three Nb atoms, such that the unit cell can be chosen to contain exactly one complete small (red) triangle, which we refer to hereafter as a trimer. 
  At low-temperature, further distortions take places~\cite{sheckelton_john_p_rearrangement_2017,haraguchi_magneticnonmagnetic_2017,kim_terahertz_2023}, reducing the symmetry, which we discuss together with the bulk structures in more detail below. 
  
  Fig.~\ref{fig:bandstructure}(a) shows the electronic band structure at the DFT level, i.e., without considering strong electronic correlations. There are three Nb $d$ $t_{2g}$ bands close to the Fermi level, of which one is crossing the Fermi level, while the other two are slightly above. At $\Gamma$ these states can be described by their $2a_1$ and $2e$ symmetries, respectively~\cite{Gao2022,haraguchi_magneticnonmagnetic_2017,sheckelton_john_p_rearrangement_2017}.
  Additional $t_{2g}$ and $e_g$ bands are above and below (green and blue). Finally, a block of Cl $p$ bands can be found below $-2.7$ eV. A Wannier construction for the $2a_1$ and $2e$ $t_{2g}$ bands provides three equivalent orbitals $\psi_{1,2,3}$ each localized at one of the three Nb atoms in the trimer, as shown in Fig.~\ref{fig:bandstructure}(c)-(e). In each case, the orbital lobes have a particular orientation with respect to the triangle of atoms, as indicated in Fig.~\ref{fig:bandstructure}(b). 
  
  An isolated trimer with only internal hopping $t_0>0$ between the three atomic Wannier orbitals has three single-particle states with energies $-2t_0$ ($2a_1$) and twice-degenerate $t_0$ ($2e$) as obtained from diagonalization of the isolated trimer Hamiltonian. Then, adding a hopping $t_\parallel \ll t_0$ between trimers leads to one low-energy band and a pair of higher energy bands, each with a bandwidth of order $t_\parallel$ and a gap of $3t_0$. 
  The strong distortion of the kagome lattice thus allows us to describe the $2a_1$ and $2e$ bands observed in Fig.~\ref{fig:bandstructure}(a) by a triangular lattice tight-binding model in a basis of \emph{molecular} (trimer) orbitals. 
  
  For an isolated trimer, the lowest molecular orbital wave function is a symmetric combination $\psi_0 = \dfrac{1}{\sqrt{3}}( \psi_1+\psi_2+\psi_3)$ of the atomic wave functions, as shown in Fig.~\ref{fig:bandstructure}(f). Correspondingly, a Wannier construction for the lowest $2a_1$ band using only one orbital per trimer yields such a molecular orbital, as shown in Fig.~\ref{fig:bandstructure}(g).
  This confirms that the trimer molecular orbital picture is a simple explanation of the low-energy single-particle electronic structure. 

  In Tables~\ref{tab:parameters:molecular} and \ref{tab:parameters:atomic}, we summarize the hopping parameters corresponding to the single molecular orbital (MO) and three-band atomic orbital (AO) Wannier constructions, respectively. The magnitudes of the hopping parameters are consistent with the picture of weakly hybridized trimers on a triangular lattice (see Methods for more details).

% Table I
\sisetup{detect-weight=true,detect-inline-weight=math}
\begin{table}[t]
    \centering
    \caption{Single molecular orbital Hubbard model parameters for monolayer Nb$_3$Cl$_8$. The upper panel depicts the triangular lattice with lattice site indices. The lower panel lists the distance $\mathbf{r}$ between the trimer lattice sites, their single-particle hopping $t$, local and non-local Coulomb interactions $U$ and effective magnetic exchange interactions $J^{(s)} = -2 (t^{(s)})^2/U^*$. All parameters are given in meV. \label{tab:parameters:molecular}}
    \begin{tabular}{
    S[table-format=2.5]
    S[table-format=2.1]
    S[table-format=4.5]
    S[table-format=4.3]
    S[table-format=3.3]
    }
        \multicolumn{5}{c}{\includegraphics[width=0.24\textwidth]{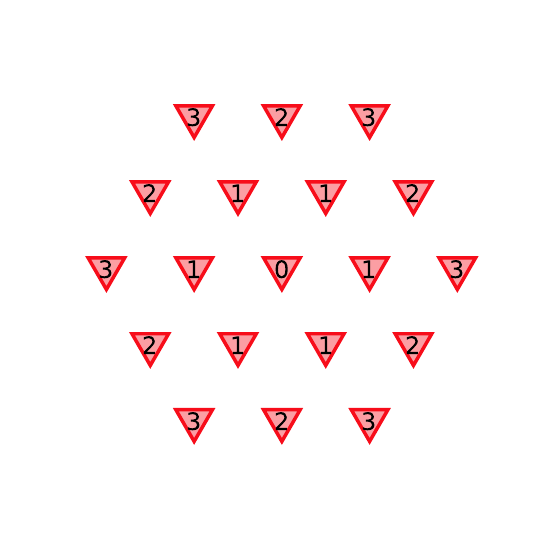}} \\
        \toprule
        {$|{\bf r}^{(s)}|/a $}   & {$s$}  & {$t^{(s)}$}  & {$U$} & {$J^{(s)}$}
                 \\ \midrule
        0.000 &0& &1907.30       &\\\hline
        1.000 &1& 22.6& 781.4& -0.91\\
        1.732 &2&  4.6& 462.0&-0.04\\
        2.000 &3& -4.0& 387.4&-0.03\\
        \bottomrule
    \end{tabular} 
\end{table}

\sisetup{detect-weight=true,detect-inline-weight=math}
\begin{table}[t]
    \centering
    \caption{Atomic Hubbard model parameters for monolayer Nb$_3$Cl$_8$. The upper panel depicts the atomic positions within the triangular lattice. The lower panel lists the distance $\mathbf{r}$ between the atomic trimer lattice sites, their single-particle hopping $t$, local and non-local density-density $U_{iijj}$ and Hund's exchange  $U_{ijji}$ Coulomb interactions. All parameters are given in meV. 
    }
    \begin{tabular}{
    S[table-format=1.1]
    S[table-format=2.3]
    S[table-format=2.3]
    S[table-format=2.]
    S[table-format=5.3]
    S[table-format=5.3]
    S[table-format=5.3]
    }
    \multicolumn{7}{c}{\includegraphics[width=0.24\textwidth]{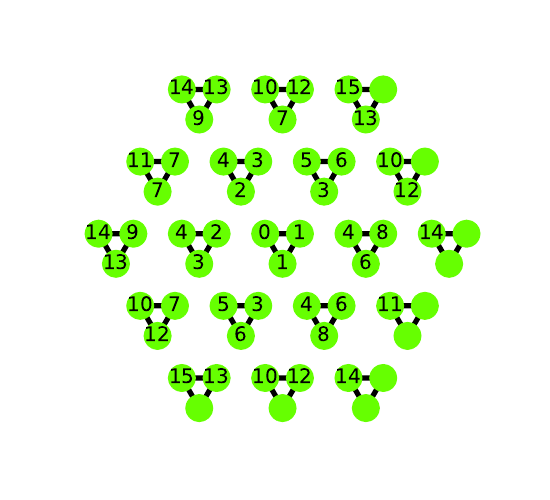}} \\
    \toprule
    {$i$-$j$} & 
    {$|{\bf r}_{ij}^{(s)}|/a $} &
    {${\bf R}_{ij}^{(s)}/a$ } & 
    {$s$} &
    {$t_{ij}^{(s)}$}  & 
    {$U_{iijj}$} & {$U_{ijji}$}
    \\ \midrule
          {1-1}  & 0.000 &  {$(000)$  }            & 0&            & \multicolumn{2}{c}{2800.2  }\\\hline
          {1-2}  & 0.417 &  {$(000)$    }          & 1& -325.1   & 1829.0 &   37.7\\
          {1-2}  & 0.583 &  {$(\bar{1}\bar{1}0)$}  & 2& 84.5     &  1018.3 &    0.2\\
          {1-2}  & 0.870 &  {$(\bar{1}00)$      }  & 3& -8.4     &  901.3 &   -0.1\\\hline
          {1-1}  & 1.000 &  {$(100)$            }  & 4& -39.8    &  \multicolumn{2}{c}{832.8}\\
          {1-1}  & 1.000 &  {$(010)$            }  & 5& 18.0     &  \multicolumn{2}{c}{852.8}\\\hline
          {1-2}  & 1.261 &  {$(100)$            }  & 6& 2.8      &  772.5 &   -0.1\\
          {1-2}  & 1.387 &  {$(\bar{1}\bar{2}0)$}  & 7& 0.4      &  561.0 &   0.0\\
          {1-2}  & 1.417 &  {$(110)$            }  & 8& -4.5     &  730.1 &   0.2\\
          {1-2}  & 1.583 &  {$(\bar{2}\bar{2}0)$}  & 9& -4.1     &  482.0 &   0.0\\\hline
          {1-1}  & 1.732 &  {$(1\bar{1}0)$      }  &10& 1.2      &  \multicolumn{2}{c}{497.8}\\
          {1-1}  & 1.732 &  {$(120)$            }  &11& -0.1     &  \multicolumn{2}{c}{501.7}\\\hline
          {1-2}  & 1.782 &  {$(1\bar{1}0)$      }  &12& 0.4      &  493.7 & 0.0\\
          {1-2}  & 1.828 &  {$(\bar{2}00)$      }  &13& -1.7     &  444.4 & 0.0\\\hline
          {1-1}  & 2.000 &  {$(\bar{2}00)$      }  &14& -1.3     &  \multicolumn{2}{c}{414.9}\\
          {1-1}  & 2.000 &  {$(200)$            }  &15& -0.5     &  \multicolumn{2}{c}{414.9}\\
        \bottomrule
    \end{tabular} 
    \label{tab:parameters:atomic}
\end{table}

\subsection{Coulomb interaction using constrained Random Phase Approximation}

  Having established two Wannier orbital basis sets, we come to our main result: the partially screened Coulomb interactions, which we calculate from first principles using the constrained Random Phase Approximation (cRPA)~\cite{Aryasetiawan04} (see Methods). To this end, the electronic structure is divided into a target space and a rest space. We calculate the effective Coulomb interaction in the target space by considering all screening processes involving rest space electrons according to the Random Phase Approximation. The cRPA is an approximation for the screening by the rest-space electrons. This approximation should be applicable when the rest space has a large gap~\cite{vanLoon21}, and the two spaces can be disentangled cleanly. Here, the symmetry of the distorted kagome lattice greatly helps the disentanglement. Furthermore, the rest space has a gap of 2.5 and 2 eV for the atomic{AO} and MO target spaces, respectively. This is sufficiently large compared to the bandwidths of the individual bands to have confidence in the cRPA accuracy. In this sense, Nb$_3$Cl$_8$ is an extraordinarily promising candidate for realizing a most simple Hubbard model.

  In the present case, there are two relevant choices for the target space: all of the three lowest $2a_1$ and $2e$ states spanned by three atomic Nb $d$ orbitals, or just the lowest metallic $2a_1$ band, best described by the lowest MO. We have performed cRPA calculations for both cases, and the resulting Coulomb matrix elements $U(R)$ are given in Tables~\ref{tab:parameters:molecular} and~\ref{tab:parameters:atomic}. For the onsite $R=0$ Coulomb interactions we find $U \approx 1.9\,$eV in the MO model and $U_{0000} \approx 2.8\,$eV ($U_{0011} \approx 1.8\,$eV) in the atomic model. $U$ and $U_{0000}$ are obviously not identical. There are two origins for these differences, one purely geometrical and one physical. The MO is constructed as a linear combination of atomic orbitals, meaning the Coulomb tensor should also be transformed to a new basis involving all intra-trimer Coulomb matrix elements yielding $U \approx 1/3 \, U_{0000} +2/3 \, U_{0011}$ (see Methods). Given the numbers in Table~\ref{tab:parameters:atomic}, more than half of the molecular Coulomb interaction comes from the interatomic interaction $U_{0011}$ within the trimer. The transformation to the MO reduces the Coulomb matrix element, since the molecular Wannier orbital, as depicted in Fig.~\ref{fig:bandstructure}(g), is more spread out than the individual atomic Wannier orbitals in Fig.~(c)-(e). In addition to this geometrical effect, the Coulomb interactions in the MO model are further reduced because of the additional screening provided by the two $2e$ bands that are integrated out. 

    \begin{figure}
        \centering
        \includegraphics[width=0.45\textwidth]{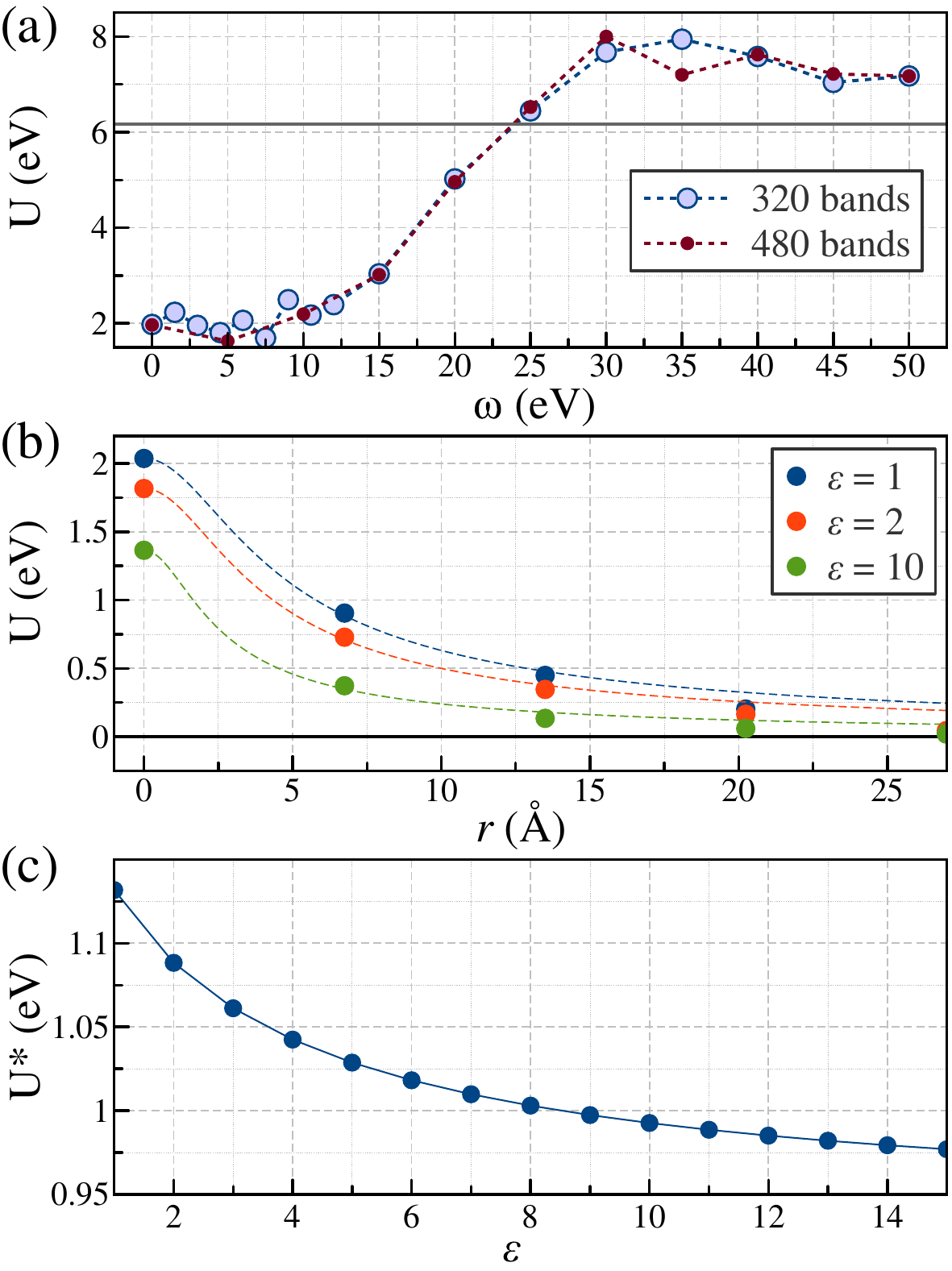}
        \caption{ 
        Coulomb interactions of the molecular single-orbital model calculated using the constrained Random Phase Approximation. (a) Frequency-dependence of the local interaction, (b) non-local static interactions for a monolayer in a dielectric environment. (c) Effective local Hubbard interaction $U^*= U(R=0) -  U(R=1)$ as a function of the dielectric environmental screening.}
        \label{fig:ur}
    \end{figure}

  Fig.~\ref{fig:ur} shows additional information about the Coulomb interaction in the MO. Formally, the cRPA produces a frequency-dependent interaction~\cite{Aryasetiawan04,Casula12} $U(\omega)$. In Nb$_3$Cl$_8$, the partially-screened Coulomb interaction is, however, well approximated as frequency-independent in the range $\omega \in [0,10]$ eV, and only reaches the asymptotic, bare value around $\omega=30$ eV. This shows that retardation effects are unimportant at the relevant electronic energy scales~\cite{Casula12}. Next to these weak retardation effects, we find significant non-local values $U(R)$ for $R>0$, as shown in Fig.~\ref{fig:ur}(b), depicting the $1/R$ tail of the static Coulomb interaction. In this figure, we also show how the non-local interaction of the molecular orbital is affected by a dielectric encapsulation of the monolayer (see Methods for details on the calculations). From this, we see that in the extrapolated fully free-standing $\varepsilon = 1$ case $U(R=0)\approx 2\,$eV and $U(R=1)\approx 0.9\,$eV, which can be reduced by an environmental screening of $\varepsilon=10$ to approx. $1.4$ and $0.4\,$eV, respectively. $U(R)$ is thus rather susceptible to the environment. However, the \emph{effective} local interaction $U^* = U(R=0)-U(R=1)$, which is the adequate quantity to use in purely local single-band Hubbard models~\cite{Schuler13}, is barely affected by the substrate screening as a result of similar screening effects to $U(R=0)$ and $U(R=1)$. This is illustrated in Fig.~\ref{fig:ur}(c), which shows that $U^*(\varepsilon)$ reduces from about $1.15$ to only $1\,$eV upon increasing the environmental dielectric screening from $\varepsilon=1$ to $10$.

\subsection{Electronic correlations}

  Solving the single-orbital triangular lattice Hubbard model is a formidable task~\cite{Wietek21} which requires advanced computational techniques and is beyond the scope of this paper. Our model derivations result in consistent parameter sets summarized in Table~\ref{tab:parameters:molecular}, which should be used for such calculations. Nevertheless, inspecting these model parameters given by our Wannier construction and cRPA calculations, we observe that the Coulomb interaction is substantially larger than the bandwidth of the partially filled low-energy band, $U/|9t^{(1)}| \approx 10$ for the molecular model. Thus, strong coupling techniques can be used to get a first impression of the correlation effects. The so-called Hubbard-I approximation, a multi-band generalization~\cite{Lichtenstein98} of the approach introduced by Hubbard~\cite{Hubbard1963}, was suggested as an adequate approximation by default in the narrow-band limit~\cite{Lichtenstein98,Locht2016} such that we will exploit it here.

    \begin{figure*}
        \includegraphics[width=0.99\textwidth]{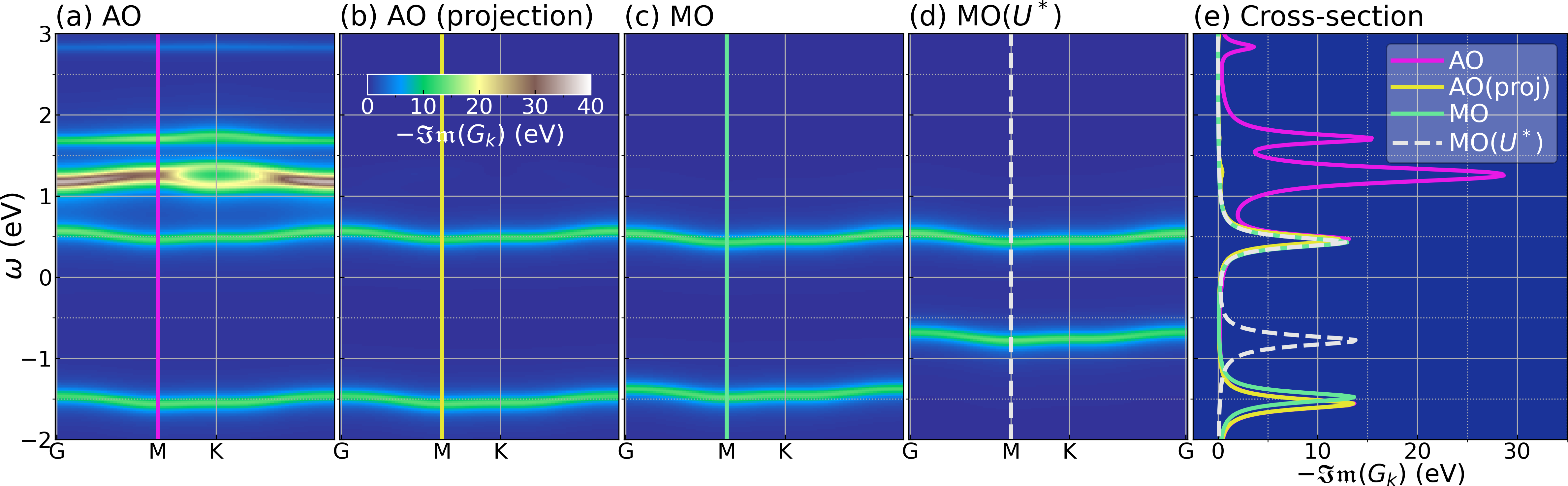}
        \caption{Interacting spectral functions of Nb$_3$Cl$_8$ monolayer within the Hubbard-I approximation. (a) Atomic three-orbital model (AO), 
        (b) its projection to the lowest molecular-orbital, 
        (c) molecular single-orbital model (MO),
        (d) MO model based on $U^*$, and
        (e) cross-section of those spectral functions at one selected momentum. The chosen chemical potential constraints the occupation to one electron per trimer and to align the upper Hubbard band.}
        \label{fig:hub1}
    \end{figure*}

  In Fig.~\ref{fig:hub1}(a) and (c), we show the spectral functions of the atomic and MO models, i.e., with 3 and 1 orbitals per trimer, calculated using the Hubbard-I approximation, taking into account the intratrimer interactions only. For the MO model, we take into account $U(R=0)$ only, while for the three-orbital model, we include the full trimer-local Coulomb tensor $U_{ijkl}$ where $i,j,k,l$ are restricted to nearest-neighbour Nb positions on a single trimer. We also show in Fig.~\ref{fig:hub1}(d) results for the MO model using $U^\ast$ to account for long-range intertrimer Coulomb interactions. In all cases, the interaction leads to a significant frequency dependence in the local Hubbard-I self-energy, which splits the partially filled $2a_1$ band into an upper and lower Hubbard band, with a gap given by the effective interaction strength acting on the lowest half-filled molecular orbital. For the atomic model, we have projected onto the trimer eigenstate $\psi_0$ to show that this is indeed the low-energy state. This comparison clearly proves that the system is a Mott insulator in the Hubbard-I approximation, regardless of the model that is used. 

  Comparing the Mott gaps in the three models, we find noticeable differences. The $U^\ast$ model obviously has a smaller gap since it uses a reduced Coulomb interaction to account for inter-trimer screening. This effect is not included at all in the approximate Hubbard-I treatment of the other models. 
  Comparing the gaps in the atomic three-orbital and the MO model, we find a small reduction of the gap in the MO model. Both approaches differ in their diagrammatic content with regard to the two higher-energy trimer states, and an exact match is, therefore, not expected. The three-orbital model in the Hubbard-I approximation contains purely intratrimer effects in an exact way. In contrast, the molecular model contains both intra- and intertrimer screening effects, but only on the (c)RPA level, while the correlation effects are subsequently taken into account using Hubbard-I, i.e., intratrimer only. The quantitative similarity is a sign that the appearance of the gap is largely caused by intratrimer correlation physics. 

  Since it starts with three bands, the atomic model obviously has additional spectral weight at higher energies as well. Fig.~\ref{fig:hub1} shows that this spectral weight predominantly remains in the tight-binding bands, with a limited amount of spectral weight transferred to localized excitations at higher energy. 

  Finally, in all cases, we observe a significant bandwidth renormalization in the lower and upper Hubbard bands as compared to the non-interacting $2a_1$ bandwidth. This is a result of the frequency dependence of the dynamical self-energy and is in agreement with recently obtained angle-resolved photo emission spectra~\cite{Gao2022}.

\subsection{Magnetic Properties}

\begin{figure}
        \centering
        \includegraphics[width=0.5\textwidth]{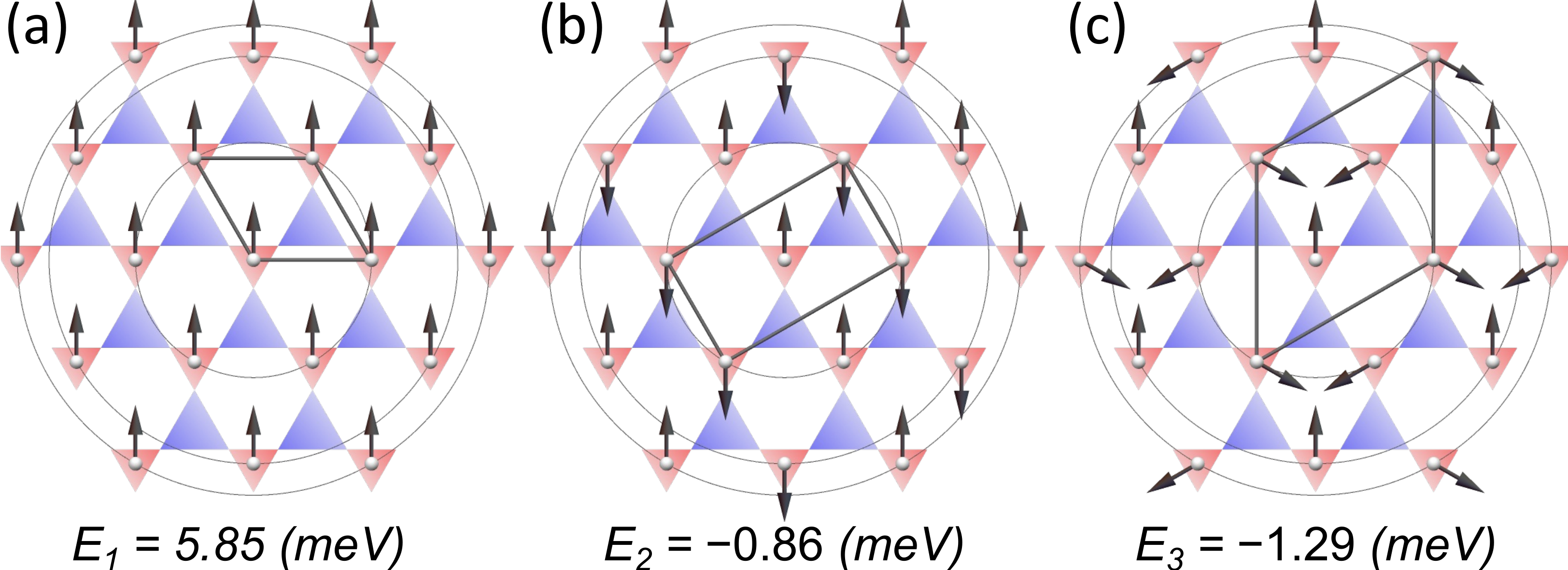}
        \caption{Magnetic structures based derived from the single-orbital Hubbard model for monolayer Nb$_3$Cl$_8$. 
        Corresponding energy densities of the exchange interaction are indicated in the Figure.}
        \label{fig:magneticstructure}
    \end{figure}

  Given the strong Coulomb repulsion and the filling of one electron per trimer, there is a tendency to form single $S=\frac{1}{2}$ magnetic moments on every trimer. In this strong coupling limit, a simple estimate for the magnetic exchange interactions is given by the Hubbard Stratonovich transformation $J^{(s)} = t^{(s)}/U^*$, and the resulting values are given in Table~\ref{tab:parameters:atomic} up to third nearest neighbors. Fig.~\ref{fig:magneticstructure} shows possible magnetic structures based on these interactions, including a ferromagnetic, striped anti-ferromagnetic, and $120^\circ$
  anti-ferromagnetic ordering together with their energy densities of the exchange interaction given by $E = -\frac{1}{2 N_i}\sum_s\sum_{ij}^{N_i N_j^{(s)}} J_{ij}^{(s)}{\bf S}_i\cdot{\bf S}_j$. Here, $i=1,\cdots, N_i$ stands for the magnetic moments in the magnetic unit cell, and $j  \neq i = 1,\cdots, N_j^{(s)}$ represent all neighbors within the same distance $|{\bf r}_{ij}^{(s)}|$ characterized by shell index $s$. From these structures, the  $120^\circ$ anti-ferromagnetic structure has the lowest energy, suggesting it adequately represents the magnetic ground state.

  Experimentally, a paramagnetic state was observed at high temperatures for bulk samples \cite{pasco_tunable_2019}. In quasi-two-dimensional magnets, magnetic order exists only at temperatures much lower than typical energy of exchange interactions due to strong thermal spin fluctuations. At intermediate temperatures, one can expect paramagnetism with strong short-range order, which can be described using, for example, self-consistent spin-wave theory \cite{irkhin99}.

\section{Discussion}

  With the help of two down-folded minimal models, we established a Hubbard model description of Nb$_3$Cl$_8$ based on correlated trimers positioned on a triangular lattice and under the influence of inter-trimer hopping as well as long-range Coulomb interactions. Based on these \textit{ab initio} models, we showed that monolayer Nb$_3$Cl$_8$ in the high-temperature (HT) phase is a Mott insulator deep in the Mott regime as a result of the integer one-electron occupation of each trimer and the strong trimer-local Coulomb interactions relative to the small inter-trimer hopping. Below we further show that this Mott benhaviour also survives in the distorted low-temperature (LT) phases as well as within bulk stacks.
  Previous studies suggested using $U\approx1$ eV as an \textit{ad hoc} choice together with a Wannierized single-band to describe bulk structures~\cite{Gao2022,Zhang23}. Our first-principles cRPA study shows that this value is indeed reasonable, once the system is mapped onto an effective Hubbard model using the renormalized $U^\ast=U(R=0)-U(R=1)$~\cite{Schuler13}, which effectively takes the non-local Coulomb interaction into account. We find slightly larger values of $U^\ast \approx 1.15$ eV for an isolated monolayer, with a reduction towards 1 eV for monolayers in a dielectric environment~\cite{roesner15,vanloon2022coulomb}. Screening by the environment is thus relatively inefficient due to the intra- and inter-trimer Coulomb interaction being both screened in approximately the same way. Based on the effective single-band Hubbard model describing ML Nb$_3$Cl$_8$, we furthermore showed that a $120^\circ$-antiferromagnetic structure forms between the  magnetic moments localized on each trimer at the ground state.

\subsection{Spin-Orbit Coupling}

  In the distorted kagome crystal structure yielding the correlated trimer lattice, Nb is formally in an [Nb$_3$]$^{8+}$ state. As the atomic spin-orbit coupling of Nb$^{2+}$ and Nb$^{3+}$ is below $100\,$meV~\cite{koseki_spinorbit_2019} and thus significantly smaller than the local Coulomb of $U_{0000} \approx 2.8\,$eV, spin-orbit coupling can be treated as a perturbation to the Mott insulating state here and will not change qualitatively the drawn conclusions. The interplay of spin-orbit coupling with the crystal field could lead however to subtleties in the magnetic anisotropy~\cite{locht_standard_2016}.

\subsection{Bulk Structures}

    \begin{figure*}
        \centering
        \includegraphics[width=0.9\textwidth]{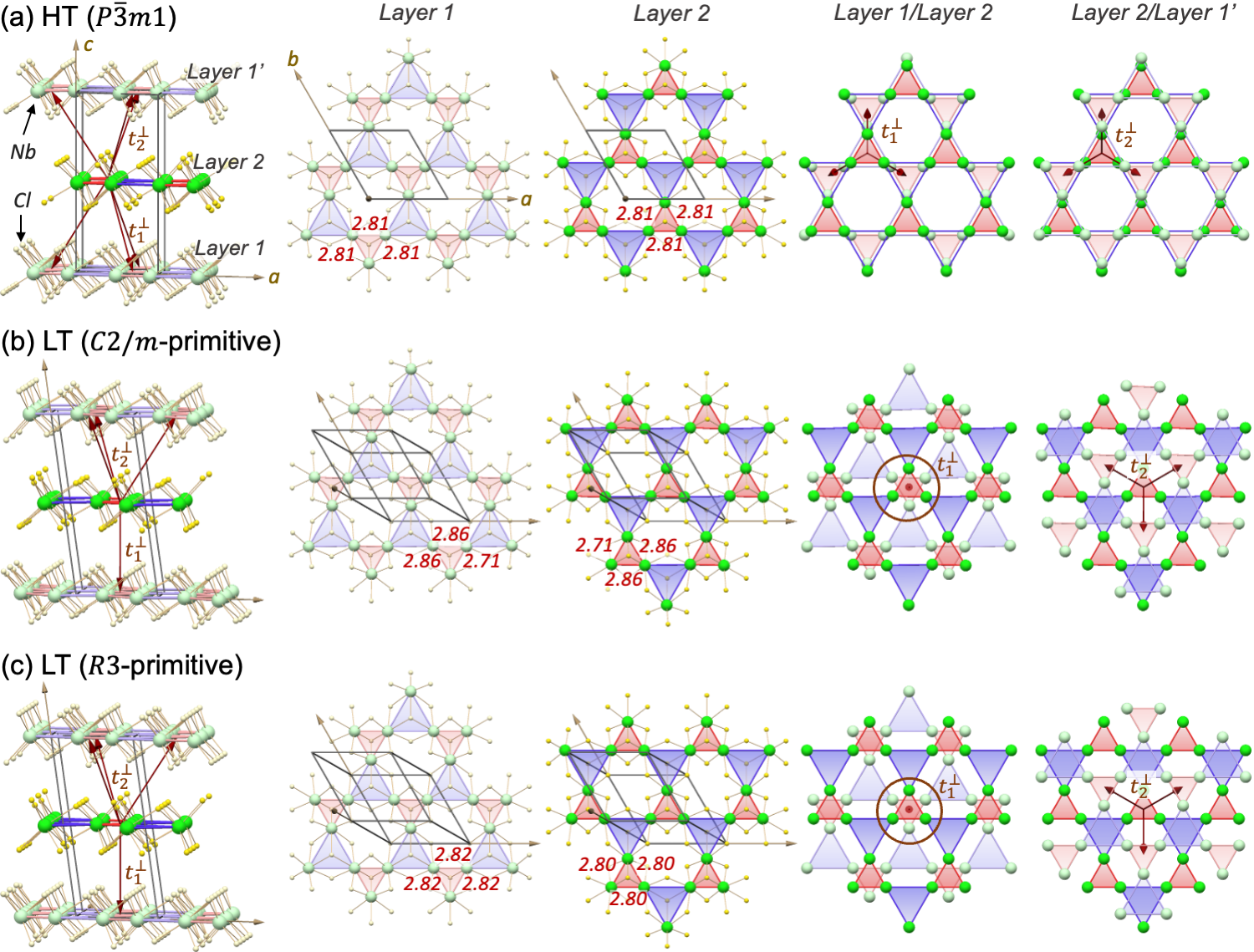}
        \caption{Crystal structures of bulk Nb$_3$Cl$_8$. (a) At high temperature (HT, space group $P\bar{3}m1$) and (b) and (c) at low temperatures (LT, space groups $C2/m$ and $R3$, respectively). The on-top views depict layer 1, layer 2, layer 1/layer 2, and layer 2/layer 1$’$, respectively. 
        Spheres in fade and bright colors (green and yellow) 
        represent atoms (of Nb and Cl) in neighboring layers.
        Brown arrows indicate the interlayer couplings $t_n^{\perp}$ ($n=1,2$) between the nearest trimers. 
        Numbers around trimers are distances between the nearest Nb atoms in each layer (in \AA).}
        \label{fig:str2}
    \end{figure*}

  In the bulk structure, the details of the stacking between the layers affect the electronic properties via interlayer hybridization. In the orthorhombic lattice structure of the HT phase, the correlated trimers in the different layers are laterally shifted as visualized in Fig.~\ref{fig:str2}(a). The resulting DFT band structure is shown in Fig.~\ref{fig:bandstructure_bulk}(a). It is similar to the ML HT band structure with a doubled amount of bands, which are mildly affected by interlayer hybridization. We now find two half-filled bands around the Fermi level with finite dispersion in $k_z$ direction. The corresponding atomic and MO models now host 6 Nb $d$ states and 2 MOs, respectively. The nearest-neighbour intra-layer hopping matrix element $t^\parallel\approx25\,$meV of the corresponding MO model is similar to the one in the ML. The nearest-neighbour inter-layer hopping $t^\perp_{1,2}\approx-15$ to $17\,$meV is on a similar order, such that the inter-layer hybridization does not qualitatively change the small-band width character of the metallic bands around the Fermi level. The direction dependence of $t^\perp_{1,2}$ is a result of the Cl positions, one of which can be either above or below the trimer.

  Next to the kinetic parts of the model Hamiltonian, the Coulomb interaction matrix elements are also affected by the bulk structure due to enhanced screening. In the MO model, the local Coulomb interaction on each trimer is reduced from $U\approx1.9\,$eV in the ML to $U\approx1.5\,$eV in the bulk structure. Similar trends hold for the long-range interactions within the individual layers. Additionally, there are now inter-layer Coulomb matrix elements, of which only the density-density ones are of non-vanishing magnitude. In the MO model, these are ca. $360\,$meV. Given the half-filled flat bands and still significant trimer-local Coulomb interaction matrix elements, it is thus reasonable to expect that the Mott behaviour in the HT bulk phase survives. The corresponding results, taking only trimer-local interaction terms into account, are presented in Fig.~\ref{fig:bandstructure_bulk} for both models. In the atomistic model, we find a Mott gap of about 
  $1.6\,$eV, and in the MO model of about $1.5\,$eV. We note that these calculations do not take non-local Coulomb interactions into account, which can further influence the Mott gap as $U^*$ did in the ML model. For the bulk structures, it is at the moment, however not clear how to construct $U^*$, since $U(R=1)$ is different along the in-plane and out-of-plane directions.

\subsection{Low-Temperature Distorted Phases}

\begin{figure*}
    \centering
    \includegraphics[width=0.95\textwidth]{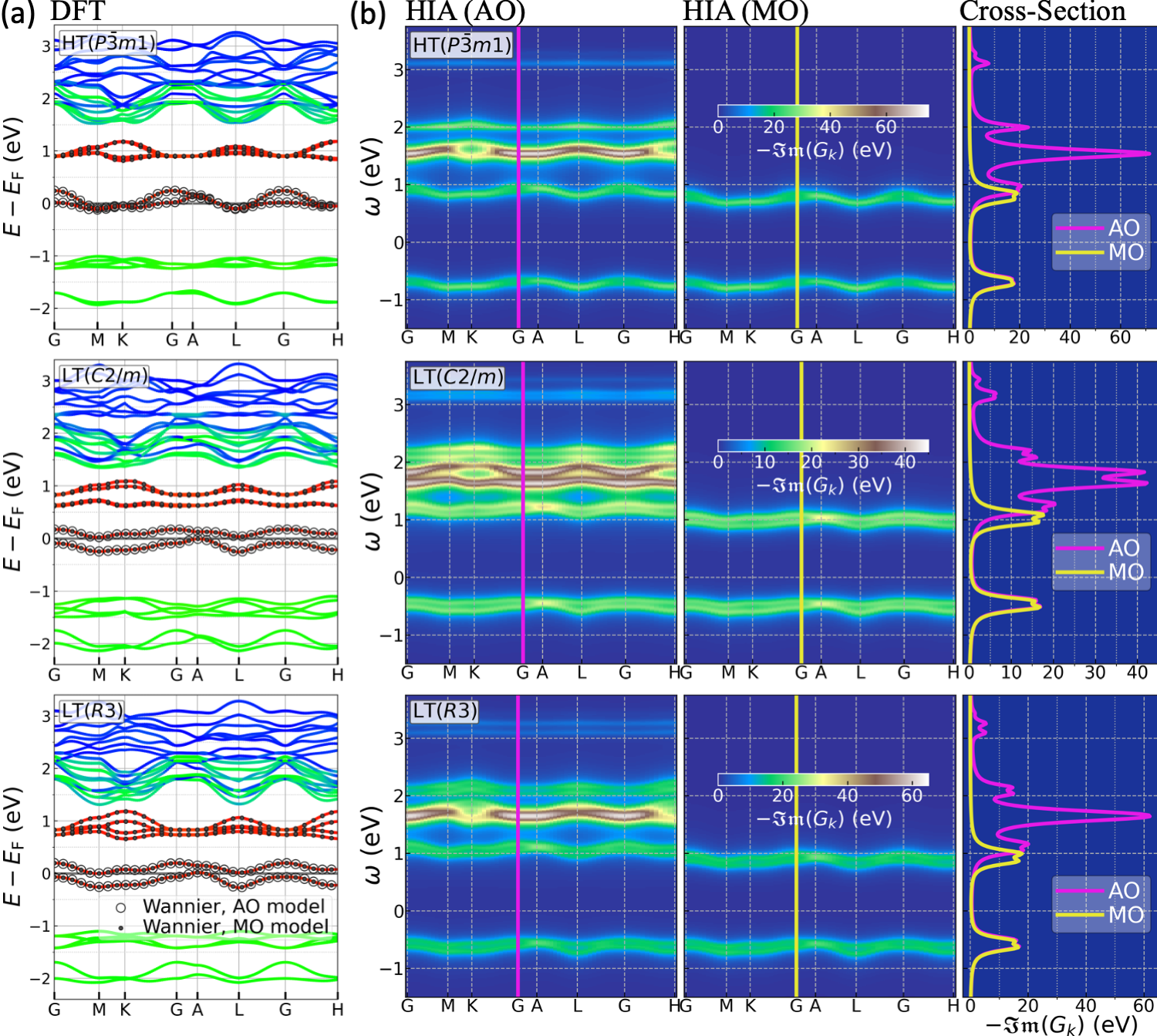}
    \caption{
    (a)	DFT electronic structure and (b) interacting spectral functions obtained using the Hubbard-I approximation for bulk Nb$_3$Cl$_8$. The top, middle, and bottom panels represent results in $P\bar{3}m1$ (HT phase), $C2/m$ (LT phase), and $R3$ (LT phase) space groups, respectively. 
    The electronic structures in (a) are weighted between Nb-$t_{2g}^1$ (in red), Nb-$t_{2g}^2$ (in green), and Nb-$e_g$ (in blue) states, respectively. 
    The spectral functions in (b) stand for the atomic six-orbital model (first panels), molecular two-orbital model (second panels), and the cross-section of those spectral functions at one selected momentum (last panels).}
    \label{fig:bandstructure_bulk}
\end{figure*}

  For low temperatures, below $100\,$K, bulk Nb$_3$Cl$_8$ is known to undergo a structural phase transition accompanied by a paramagnetic to non-magnetic transition~\cite{haraguchi_magneticnonmagnetic_2017,sheckelton_john_p_rearrangement_2017,pasco_tunable_2019,kim_terahertz_2023}. Two mildly different low-temperature structures with $C2/m$~\cite{sheckelton_john_p_rearrangement_2017} and $R3$~\cite{haraguchi_magneticnonmagnetic_2017,kim_terahertz_2023} point group symmetry have been experimentally observed. The corresponding DFT band structures are shown in Fig.~\ref{fig:bandstructure_bulk}.
  The main difference between the $C2/m$ and $R3$ structures is a deformation within the trimers, yielding slightly elongated trimers in the $C2/m$ structure, and modifications to the trimer sizes between the different layers in the $R3$ structure, as depicted in Fig.~\ref{fig:str2}. As a result, the degeneracies between the two higher-lying trimer states are lifted in both LT structures.
  The main difference between these LT and HT bulk structures is, however, the relative shift between the layers. In the LT structures, trimers from different layers are partially significantly closer to each other than in the HT structure, c.f. Fig.~\ref{fig:str2}. This enhances the inter-layer hybridization, which leads to a splitting of the two metallic bands around the Fermi level, which yields a full but small gap between these states. On the DFT level, it is important to note that none of the symmetry breakings in the $C2/m$ and $R3$ structures yield significant charge re-distributions between the Nb atoms of the trimers.

  The molecular orbital description is furthermore still valid and allows to quantify of the nearest-neighbor intra- and inter-layer hoppings yielding $t^\parallel\approx\,15$meV, $t_1^\perp\approx127\,$meV, $t_2^\perp\approx 3\,$meV, respectively for the $R3$ structure. Notably, $t_2^\perp$ is much smaller than $t_1^\perp$, which is in turn much larger than $t^\parallel$. The LT structures is thus best described as weakly hybridized stacks of bilayer Nb$_3$Cl$_8$, whereby the two layers forming the bilayers are strongly hybridized. The Coulomb interaction matrix elements are only barely affected by the relative shift between the two layers of the LT bulk unit cell yielding $U\approx 1.5\,$eV in the $R3$ structures, respectively. A similar description holds for the $C2/m$ structure, however, with reduced symmetry. All model parameters are given in 
  Table~\ref{tab:parameters_bulk}.

  We have also applied the Hubbard-I approximation to the bulk structures. 
  As shown in Ref.~\cite{Svane2005}, the Hubbard-I approach corresponds to the formal first-order-in hopping correction to the Green's function, assuming that the interaction is local, which leads to vanishing vertex corrections. This assumption is still clearly valid in the case of the LT structures as the interactions between the trimers, as well as the hopping between them, is much smaller than the intertrimer Coulomb interaction. The only other factor which could yield vertex corrections is correlated hopping~\cite{polar1934,VKI1979,Tom22}. Explicit calculations show here, however, that these contributions to the hopping are small: The largest $U_{iiij}$ element with $i$ and $j$ being located on different trimers, which could be responsible for correlated hopping terms, is on the order of just $3\,$meV and can thus be safely neglected. In Fig.~\ref{fig:bandstructure_bulk}, we, therefore, present the corresponding interacting spectral functions for both atomic and MO models. Similar to the situations studied above, we find Mott gaps on the order of $1.6\,$eV driven by the dynamical properties of the Hubbard-I self-energy. Due to the small splitting of the two bands around the Fermi level, we see a similarly small splitting in the lower and upper Hubbard bands as well. We again note, that these calculations do not include the effects of non-local Coulomb interaction terms.

  It is worthwhile to note that the LDA+$U$ approximation, popular due to its simplicity, is much less accurate than the Hubbard-I approximation for narrow-band systems such as Nb$_3$Cl$_8$. As discussed in detail for elemental rare-earth and their compounds~\cite{Peters2014,Locht2016}, Hartree-Fock calculations utilized within LDA+U~\cite{Hubbard1963} cannot account for the significant bandwidth renormalizations, which are captured by Hubbard-I calculations and which have been observed experimentally in Nb$_3$Cl$_8$~\cite{Gao2022,sun_observation_2022}.
   
\section{Conclusions}

  Based on detailed \textit{ab initio} down folding calculations, we were able to derive various minimal (generalized) Hubbard models for Nb$_3$Cl$_8$ in several structures. For all of them, the derived model parameters, together with the Hubbard-I approximation suggest that Nb$_3$Cl$_8$ is a Mott insulator, independent of the details of the low- or high-temperature bulk or monolayer structures. 
  For the monolayer structures, we were able to derive the most simple single-band Hubbard model. For bulk structures in all known lattice structures, we derived two-orbital Hubbard models with long-range Coulomb interactions.

  The sister compound Nb$_3$Br$_8$ has been suggested to be an ``obstructed atomic insulator''~\cite{xu2021threedimensional} already at the DFT level, based on symmetry analysis. Our results for Nb$_3$Cl$_8$ suggest that correlation effects are also important in Nb$_3$Br$_8$ and that these would give a substantial contribution to any experimentally observed gap. To quantify this, cRPA calculations for Nb$_3$Br$_8$ are needed.

  Non-local Coulomb interactions are substantial in this layered compound. For the monolayer, it is possible to use an effective Hubbard model with a modified Hubbard interaction to account for this~\cite{Schuler13}. We need to stress that the role of non-local Coulomb interactions in the bulk structure should be investigated in more detail, since the effective Hubbard model approach assumes that the non-local Coulomb interaction is the same in all directions, which is not the case in this layered compound.
  
  The analysis of electronic correlations in this work (using Hubbard-I) is restricted to the undoped case, with one electron per trimer. Doping away from this integer filling will destroy the Mott insulating phase and thus requires a more advanced many-body treatment of the arising correlation phenomena. The model parameters derived here make this possible and thus form a promising basis to study the iconic Hubbard model and its physics hand-in-hand between theory and experiment.

\section{Methods}

\subsection{Electronic structure calculations}

\subsubsection{Cyrstal structures}
\label{str_data}

\sisetup{detect-weight=true,detect-inline-weight=math}
\begin{table}[t]
    \centering
    \caption{Wyckoff positions for bulk Nb$_3$Cl$_8$ characterized by $P\bar{3}m1$, $R3$, and $C2/m$ space groups. The data is taking from Refs.~\cite{haraguchi_magneticnonmagnetic_2017,sheckelton_john_p_rearrangement_2017} and  modified using allowed symmetry operations in such a way that the transition from HT phase ($P\bar{3}m1$) to LT phase ($R3$ or $C2/m$) is clear, as demonstrated in Fig.~\ref{fig:str2}.}
    \begin{tabular}{
    S[table-format=2.]
    S[table-format=1.]
    S[table-format=3.6]
    S[table-format=3.6]
    S[table-format=3.6]
    }
        \toprule
        {atom}   &  {site} & {$x$} & {$y$} & {$z$}\\ \midrule
        \multicolumn{5}{c}{HT--$P\bar{3}m1$} \\\cmidrule(lr){1-5}
        {Nb1} & {6i} & 0.527625 & 0.055250 & 0.000000\\
        {Cl1} & {2d} & 0.666667 & 0.333333 & 0.852030 \\
        {Cl2} & {2d} & 0.333333 & 0.666667 & 0.109730 \\
        {Cl3} & {6i} & 0.669600 & 0.834800 & 0.889230 \\
        {Cl4} & {6i} & 0.337400 & 0.168700 & 0.137730 \\\midrule
        \multicolumn{5}{c}{LT--$R3$} \\\cmidrule(lr){1-5}
        {Nb1} & {9b} & 0.693833 & 0.887667 & 0.000000\\
        {Nb2} & {9b} & 0.305230 & 0.110400 & 0.169523\\
        {Cl1} & {3a} & 0.500000 & 0.500000 & 0.038670\\
        {Cl2} & {9b} & 0.329633 & 0.670367 & 0.962490\\
        {Cl3} & {9b} & 0.164600 & 0.332400 & 0.205223\\
        {Cl4} & {3a} & 0.833333 & 0.166667 & 0.950790\\
        {Cl5} & {3a} & 0.166667 & 0.833333 & 0.218773\\
        {Cl6} & {9b} & 0.496700 & 0.998100 & 0.124123\\
        {Cl7} & {9b} & 0.000433 & 0.999867 & 0.043830\\
        {Cl8} & {3a} & 0.500000 & 0.500000 & 0.134162\\\midrule
        \multicolumn{5}{c}{LT--$C2/m$} \\\cmidrule(lr){1-5}
        {Nb1} & {4i} & 0.812468 & 0.000000 & 0.502000\\
        {Nb2} & {8j} & 0.099417 & 0.702000 & 0.510000\\
        {Cl1} & {4i} & 0.462684 & 0.000000 & 0.378100\\
        {Cl2} & {4i} & 0.967717 & 0.000000 & 0.402300\\
        {Cl3} & {4i} & 0.725650 & 0.000000 & 0.656300\\
        {Cl4} & {4i} & 0.205050 & 0.000000 & 0.620600\\
        {Cl5} & {8j} & 0.206317 & 0.249000 & 0.375900\\
        {Cl6} & {8j} & 0.454183 & 0.239400 & 0.617200\\
        \bottomrule
    \end{tabular} 
    \label{tab:str}
\end{table}

We study the electronic and magnetic properties of Nb$_3$Cl$_8$ using the unit cells of $P\bar{3}m1$, $R3$, and $C2/m$ space groups~\cite{haraguchi_magneticnonmagnetic_2017,sheckelton_john_p_rearrangement_2017}. These unit cells have the following bases (unit cell vectors):
$$
\mat{\mathcal{A}}^{P\bar{3}m1} = 
\frac{1}{2}
\begin{pmatrix}
\ph{-} \sqrt{3}a_1 & a_1 & 0 \\
-\sqrt{3}a_1 & a_1 & 0 \\
0 & 0 & 2 c_1
\end{pmatrix},
$$
with $a_1 = 6.74566$~\AA\, and $c_1=12.28056$~\AA,
$$
\mat{\mathcal{A}}^{R3} = 
\frac{1}{2}
\begin{pmatrix}
\ph{-} \sqrt{3}a_2 & a_2 & 0 \\
-\sqrt{3}a_2 & a_2 & 0 \\
0 & 0 & 2 c_2
\end{pmatrix},
$$
with $a_2=6.74566$~\AA\, and $c_2=36.75$~\AA
$$
\mat{\mathcal{A}}^{C2/m} = 
\begin{pmatrix}
a_3 & 0 & 0 \\
 0 & b_3 & 0 \\
c_3 \cos \beta_3 & 0 & c_3 \sin \beta_3
\end{pmatrix},
$$
with
$a_3=11.6576$\AA, $b_3=6.7261$\AA, 
$c_3=12.8452$\AA,
and $\beta_3 = 90^\circ + \arcsin(a_3/3c_3) = 107.6087^\circ$.
While $P\bar{3}m1$ is a primitive unit cell 
and contains only 22 atoms, $R3$ and $C2/m$ are conventional, containing 66 and 44 atoms, respectively. 
The primitive representations of  unit cells with $R3$ and $C2/m$ symmetries have the following forms, 
respectively:
$$
\mat{\mathcal{A}}^\text{R3} \rightarrow
\mat{\mathcal{A}}^\text{R3(p)} = 
\frac{1}{2}
\begin{pmatrix}
\ph{-} \sqrt{3}a_2 & a_2 & 0 \\
-\sqrt{3}a_2 & a_2 & 0 \\
-2 a_2 /\sqrt{3} & 0 &  2c_2/3
\end{pmatrix},
$$
$$
\mat{\mathcal{A}}^{C2/m} \rightarrow
\mat{\mathcal{A}}^\text{C2/m(p)} = 
\begin{pmatrix}
\ph{-}a_3/2 & b_3/2 & 0 \\
-a_3/2 & b_3/2 & 0 \\
c_3 \cos \beta_3 & 0 & c_3 \sin \beta_3
\end{pmatrix},
$$
Crystal structures with primitive bases $\mat{\mathcal{A}}^\text{R3(p)}$ and $\mat{\mathcal{A}}^\text{C2/m(p)}$,
similarly to the structure with basis $\mat{\mathcal{A}}^{P\bar{3}m1}$, 
contain only 22 atoms. 
The  atomic  positions for the unit cells in $P\bar{3}m1$, $R3$, and $C2/m$ space groups are taken from  Refs.~\cite{haraguchi_magneticnonmagnetic_2017,sheckelton_john_p_rearrangement_2017} and
their values in a new basis are obtained as ${\bf r}^\text{new}_i = {\bf r}^\text{old}_i\cdot \mat{\mathcal{A}}^\text{old}\cdot (\mat{\mathcal{A}}^\text{new})^{-1}$. 
To get the electronic structures in terms of pure $t_{2g}$ and $e_g$ states for a triangular lattice, the above basis vectors were rotated such that all lattice triangle sides in each layer form the cube face diagonals in a rotated frame, see Fig.~\ref{fig:bandstructure}(b).
Monolayers are constructed from these bulk structures.

    \subsubsection{Wannier Basis Sets and Non-Interacting Hamiltonians}

        We utilize two different Wannier basis sets representing the lowest trimer molecular orbital alone and an atomistic one describing the three relevant Nb $d$ orbitals on the trimer-corner positions, which we refer to as the molecular orbital (MO) and atomic orbial (AO) models, respectively. To this end, we start with conventional DFT calculations utilizing the Perdew–Burke–Ernzerhof GGA exchange-correlation functional \cite{PhysRevLett.77.3865} within a PAW basis \cite{paw,paw2} as implemented in the \emph {Vienna Ab initio Simulation Package} ({\sc vasp}) \cite{KRESSE199615,PhysRevB.54.11169} using the structures given above. 
        We use ($20 \times 20 \times 1$) and ($8 \times 8 \times 4$) $k$-meshes for monolayer and bulk calculations, respectively, as well as an energy cutoff of $350\,$eV and Methfessel-Paxton smearing of $\sigma=0.05\,$eV. 
        
        For the MO model, we project the flat Kohn-Sham bands around the Fermi level to an (initial) Wannier orbital with $s$ symmetry located at the center of the trimer. For the atomistic model, we similarly project all Kohn-Sham states between $-1.0\,$eV and $+1.5\,$eV to three individual Nb-centered $d$ orbitals. Afterward, we maximally localize the orbitals using the {\sc wannier90} package \cite{MOSTOFI2008685} and applying an inner (frozen) window including either only the flat bands around $E_F$ or all states between $-1.0\,$eV and $+1.5\,$eV. Using the resulting localized orbitals $\psi_i(r)$, we obtain the single-particle Wannier Hamiltonian from the hopping matrix elements
        \begin{align}
            t_{ij} = -\langle \psi_i \vert H_{\text{DFT}} \vert \psi_j \rangle.
        \end{align}
        The resulting Wannier models perfectly interpolate all Kohn-Sham states in their respective windows, as indicated in Fig.~\ref{fig:bandstructure} and Fig.~\ref{fig:bandstructure_bulk}.

\sisetup{detect-weight=true,detect-inline-weight=math}
\begin{table*}[t]
    \centering
    \caption{Generalized Hubbard model parameters for bulk Nb$_3$Cl$_8$ in HT and LT phases for the two molecular orbital models. Tables list in-plane $t_{ij}^{\parallel (s)}$ and out-of-plane $t_{ij}^{\perp (s)}$ hopping parameters together with local and non-local Coulomb matrix elements. $i$ and $j$ are molecular orbital positions separated by distance  $|{\bf r}_{ij}^s|/a = |({\bf r}_j + {\bf R}_{ij}) - {\bf r}_i|/a$ (in units of the in-plane lattice parameter $a$), where ${\bf R}_{ij}$ is a unit cell translation vector, and $s$ is shell index (represents bonds with the same symmetries).
}
\begin{tabular}{
S[table-format=1.]
S[table-format=1.3]
S[table-format=3.1]
S[table-format=2.2]
S[table-format=5.3]
S[table-format=5.3]
}
\toprule
\multicolumn{6}{c}{HT--$P\bar{3}m1$} \\\cmidrule(lr){1-6}
{$s^\parallel$}   &  {$|{\bf r}_{ij}^{\parallel(s)}|/a $} & {$i$-$j$} & {${\bf R}_{ij}^{\parallel(s)}/a$ } & {$t_{ij}^{\parallel(s)}$} & {$U_{ij}^{\parallel(s)}$}\\ \midrule
 0    &    0.000    &    1 1    &    {$(       0       0       0)$}    &     4037.2  &  1477.6\\ \midrule
 1    &    1.000    &    1 1    &    {$(\bar{1}\bar{1}       0)$}      &       25.4  &   474.2\\ 
 2    &    1.732    &    1 1    &    {$(\bar{2}\bar{1}       0)$}      &        4.6  &   278.9\\
 3    &    2.000    &    1 1    &    {$(\bar{2}\bar{2}       0)$}      &       -4.2  &   245.7\\
\cmidrule(lr){1-6}
{$s^\perp$}   &  {$|{\bf r}_{ij}^{\perp(s)}|/a $} & {$i$-$j$} & {${\bf R}_{ij}^{\perp(s)}/a$ } & {$t_{ij}^{\perp(s)}$} & {$U_{ij}^{\perp(s)}$}\\ \midrule
 1    &    1.046    &    1 2    &    {$(      0  0       0)$}     &      -15.5  &  358.0 \\
 2    &    1.110    &    1 2    &    {$(       0\bar{1}\bar{1})$}      &      -16.8  &   319.2\\
 3    &    1.447    &    1 2    &    {$(       1\bar{1}       0)$}     &       -0.5  &   277.8\\
 4    &    1.494    &    1 2    &    {$(\bar{1}\bar{1}\bar{1})$}       &       -4.4  &   261.6\\
 5    &    1.759    &    1 2    &    {$(\bar{1}\bar{2}       0)$}      &        1.1  & 244.6\\
 6    &    1.798    &    1 2    &    {$(       0\bar{2}\bar{1})$}      &       -0.9  & 231.0\\%\hline
 7    &    1.821    &    1 1    &    {$(       0       0\bar{1})$}     &       -1.1  & 193.2\\%\hline
 8    &    2.077    &    1 1    &    {$(\bar{1}\bar{1}\bar{1})$}       &       -0.3  & 179.9\\
 \bottomrule
\end{tabular} \\
\begin{tabular}{
S[table-format=1.]
S[table-format=1.3]
S[table-format=3.1]
S[table-format=2.2]
S[table-format=5.3]
S[table-format=5.3]
}
\toprule
\multicolumn{5}{c}{LT--$C2/m$} \\\cmidrule(lr){1-6}
{$s^\parallel$}   &  {$|{\bf r}_{ij}^{\parallel(s)}|/a $} & {$i$-$j$} & {${\bf R}_{ij}^{\parallel(s)}/a$ } & {$t_{ij}^{\parallel(s)}$}& {$U_{ij}^{\parallel(s)}$}\\ \midrule
 0    &    0.000    &    1 1    &    {$(      0      0      0)$}    &     4285.0  & 1476.1\\ \midrule
 1    &    1.000    &    1 1    &    {$(\bar{1}\bar{1}      0)$}    &       7.7  & 481.7\\
 2    &    1.000    &    1 1    &    {$(\bar{1}      0      0)$}    &       22.7  & 481.2\\
 3    &    1.731    &    1 1    &    {$(\bar{2}\bar{1}      0)$}    &       5.4  & 286.3\\
 4    &    1.732    &    1 1    &    {$(\bar{1}       1       0)$}  &       5.2  & 288.2\\
 5    &    2.000    &    1 1    &    {$(\bar{2}       0       0)$}  &      -4.4  & 253.8\\
\cmidrule(lr){1-6}
{$s^\perp$}   &  {$|{\bf r}_{ij}^{\perp(s)}|/a $} & {$i$-$j$} & {${\bf R}_{ij}^{\perp(s)}/a$ } & {$t_{ij}^{\perp(s)}$}& {$U_{ij}^{\perp(s)}$}\\ \midrule
 1    &    0.860    &    1 2    &    {$(      0      0      0)$}    &     -133.9  & 409.4\\
 2    &    1.113    &    1 2    &    {$(\bar{1}      0\bar{1})$}    &       -2.2  & 316.0\\
 3    &    1.133    &    1 2    &    {$(      0      0\bar{1})$}    &       -3.9  & 316.8\\
 4    &    1.302    &    1 2    &    {$(\bar{1}      0      0)$}    &        3.1  & 285.2\\
 5    &    1.319    &    1 2    &    {$(\bar{1}\bar{1}      0)$}    &        5.4  & 283.2\\
 6    &    1.336    &    1 2    &    {$(      0\bar{1}      0)$}    &        7.2  & 284.0\\
 7    &    1.482    &    1 2    &    {$(\bar{1}      1\bar{1})$}    &       -6.1  & 264.1\\
 8    &    1.511    &    1 2    &    {$(\bar{1}\bar{1}\bar{1})$}    &       -5.6  & 263.4\\
 9    &    1.787    &  {1-2}    &  {$(      0      2\bar{1})$}      &    -1.5  &   233.9 \\
10    &    1.799    &  {1-2}    &  {$(      1      2\bar{1})$}      &    -0.9  &   233.4 \\
11    &    1.824  &  {1-2}      &  {$(      0\bar{1}\bar{1})$}      &    -1.1  &   234.4 \\
\bottomrule
\end{tabular} \quad
\begin{tabular}{
S[table-format=1.]
S[table-format=1.3]
S[table-format=3.1]
S[table-format=2.2]
S[table-format=5.3]
S[table-format=5.3]
}
\toprule
\multicolumn{5}{c}{LT--$R3$} \\\cmidrule(lr){1-6}
{$s^\parallel$}   &  {$|{\bf r}_{ij}^{\parallel(s)}|/a $} & {$i$-$j$} & {${\bf R}_{ij}^{\parallel(s)}/a$ } & {$t_{ij}^{\parallel(s)}$} & {$U_{ij}^{\parallel(s)}$}\\ \midrule
 0    &    0.000    &    1 1    &    {$(      0      0      0)$}    &     4164.4  & 1498.0\\ 
 1    &    0.000    &   2 2    &    {$(      0      0      0)$}     &     4159.9  & 1494.0\\ 
 \midrule
 2    &    1.000    &    1 1    &    {$(\bar{1}\bar{1}      0)$}    &       15.1  & 484.2\\
 3    &    1.000    &    2 2    &    {$(\bar{1}\bar{1}      0)$}    &       26.1  & 486.1\\
 4    &    1.732    &    1 1    &    {$(\bar{2}\bar{1}      0)$}    &        4.8  & 288.6\\
 5    &    1.732    &    2 2    &  {$(      1      2      0)$}      &        5.5  &   288.5 \\
 %\\
 6    &    2.000    &    1 1    &    {$(\bar{2}\bar{2}       0)$}   &    -4.8  & 254.9\\
 7    &    2.000    &    2 2    &  {$(      0      2      0)$}      &    -4.0  &  254.7\\
\cmidrule(lr){1-6}
{$s^\perp$}   &  {$|{\bf r}_{ij}^{\perp(s)}|/a $} & {$i$-$j$} & {${\bf R}_{ij}^{\perp(s)}/a$ } & {$t_{ij}^{\perp(s)}$}& {$U_{ij}^{\perp(s)}$}\\ \midrule
 1    &    0.870    &    1 2    &    {$(      0      0      0)$}    &     -127.4  & 409.8\\
 2    &    1.108    &    1 2    &    {$(      0      0\bar{1})$}    &       -3.0  & 323.5\\
 3    &    1.325    &    1 2    &    {$(      0\bar{1}      0)$}    &        5.6  & 287.1\\
 4    &    1.492    &    1 2    &    {$(\bar{1}\bar{1}\bar{1})$}    &       -6.5  & 268.6\\
 5    &    1.796    &    1 2    &  {$(      1      2\bar{1})$}      &    -1.2  &   237.5 \\
 6    &    1.906    &    1 1    &  {$(      0      0      1)$}      &    -1.1  &   182.2 \\
 7    &    1.938    &    1 2    &  {$(      1      2      0)$}      &     0.9  &   221.7 \\
 8    &    1.938    &    1 2    &  {$(\bar{1}      1      0)$}      &    -1.3  &   221.9 \\
 9    &    2.181    &    1 2    &  {$(      2      2      0)$}      &     0.6  &   205.5 \\
%10    &    2.286    &    1 2    &  {$(      2      2\bar{1})$}      &     0.0  &   202.2 \\
%11    &    2.495    &    1 2    &  {$(      1      3\bar{1})$}      &    -0.1  &   190.9 \\
 \bottomrule
 \end{tabular} 
 \label{tab:parameters_bulk}
\end{table*}

    \subsubsection{Coulomb Matrix Elements from Constrained Random Phase Approximation Calculations}

        The Coulomb interaction matrix elements $U_{ijkl}$ are evaluated within the Wannier orbital basis sets from above and using the constrained random phase approximation (cRPA) scheme \cite{Aryasetiawan04} according to
        \begin{align}
            U_{ijkl}(\omega)&= \braket{\psi_i  \psi_k \vert \widehat{U}(\omega) \vert \psi_l  \psi_j } \\
                    &= \int \int d^3r \, d^3r' \, \psi_i^*(r) \psi_j(r) \, U(r,r') \, \psi^*_k(r') \psi_l(r') \notag
        \end{align}
        which utilizes the partially screened Coulomb interaction
        \begin{align}
        \label{eq:screen_U}
            \widehat{U}(\omega) = \left[ 1- \hat{v} \, \widehat{\Pi}_\text{cRPA}(\omega) \right]^{-1} \hat{v}.
        \end{align}
        Here $\hat{v}$ is the bare Coulomb interaction and $\widehat{\Pi}_\text{cRPA}$ is the partial polarization as defined
        \begin{align}
            \widehat{\Pi}_\text{cRPA} = \widehat{\Pi}_\text{full} - \widehat{\Pi}_\text{target},
        \end{align}
        with $\Pi_\text{target}$ describing all polarization contributions from within the target sub-space as defined by the Wannier functions. This way, we avoid any double counting of screening channels in subsequent solutions of the derived generalized Hubbard models.
        ${\Pi}_\text{cRPA}$ correspondingly describes screening from all other states except the target-ones including from a significant amount of empty states from the full Kohn-Sham basis. In detail, we use in total $160$ bands and define ${\Pi}_\text{cRPA}$ using Kaltak's projector method as recently implemented in {\sc vasp} \cite{KaltakcRPA}.

        From this we derive the full retarded and non-local rank-4 Coulomb tensor $U_{ijkl}(\omega)$ for both models.
        Casula \emph{et~al}. showed that using $U(\omega=0)$ instead of the fully retarded $U(\omega)$ is justified when renormalized hopping parameters are utilized~\cite{PhysRevLett.109.126408}. The corresponding renormalization factor was shown to scale with the characteristic cRPA plasmon frequency $\omega_\text{p}$.
        To this end, we show in Fig.~\ref{fig:ur} $U(\omega)$ of the MO orbital model in the HT ML structure. This indicates that $\omega_\text{p} \approx 25\,$eV is large such that we can safely neglect retardation effects here and use only $U_{ijkl}(\omega=0)$.

        All monolayer calculations are performed within a supercell of $25\,$\AA\ height. All tabulated and mentioned Coulomb matrix elements refer to these calculations. To extrapolate to the effectively free-standing monolayer and to take dielectric screening from the environment into account, we use our Wannier Function Continuum Electrostatics (WFCE) approach~\cite{roesner15} as applied in Ref.~\cite{soriano_environmental_2021}.

\subsection{Trimer model}

A single trimer can be described in tight-binding theory using the Hamiltonian $\hat{H}= t_0 \begin{pmatrix} 0 & 1 & 1 \\ 1 & 0 & 1 \\ 1 & 1 & 0 \end{pmatrix}$, which has eigenvalues $E_0=2t_0$, $E_\pm = -t_0$ (note that $t_0$ is negative, so $E_0<E_\pm$) and corresponding eigenvectors
\begin{align}
    \psi_0 = \frac{1}{\sqrt{3}}
    \begin{pmatrix}
    1 \\ 1 \\ 1    
    \end{pmatrix},
    \psi_+ = \frac{1}{\sqrt{3}}
    \begin{pmatrix}
    1 \\ \lambda \\ \lambda^2
    \end{pmatrix},    
    \psi_- = \frac{1}{\sqrt{3}}
    \begin{pmatrix}
    1 \\ \lambda^{-1} \\ \lambda^{-2}
    \end{pmatrix},    
\end{align}
where $\psi_1$ are the atomic orbital wavefunctions and $\lambda=\exp(2\pi i/3)$. For the Nb$_3$Cl$_8$ trimer, $t_0=-0.325$ eV, so the trimer gap $3 |t_0| \approx 1$ eV.

In the lattice of trimers, we have to take into account the hopping between trimers (i.e., along the blue bonds in Fig.~\ref{fig:latticestructure}). In the molecular orbital model, the hopping between trimers is found to be $t_1\approx 22$ meV. In the atomic orbital model, the hopping from the corner of one trimer to the nearest atom on the neighboring trimer has an amplitude of $84$ meV. Based only on this, one might have expected $t_1 \approx 84/3=28$ meV, but sub-leading matrix elements (with alternating signs) are responsible for corrections. For a triangular lattice with nearest-neighbor hopping, the bandwidth is $9t_1 \approx 200$ meV. 

For the Coulomb interaction, there are two effects that play a role when going from the atomic orbitals to the molecular orbitals. First, there is a purely geometric effect, since the molecular orbitals are linear combinations of the atomic orbitals, which leads to a transformation of the Coulomb tensor. Secondly, additional screening processes take place when integrating out the two higher-lying molecular orbitals $\psi_\pm$, which changes the concept of partially screened. Here, we consider the first effect, since it can be understood in simple terms. In this calculation of the Coulomb interaction, it is important to realize that changes in the single-particle energies (chemical potential) occur simultaneously with the transformation of the Coulomb matrix. Therefore, the best way to relate matrix elements of two different Hamiltonians is to evaluate energy differences between sets of states with constant total particle number. In this case, we look at the energy of the trimer for $n_\uparrow,n_\downarrow=0,1$. We use the fact that the electron density is homogeneously distributed over the three atoms for the low-lying molecular orbital, so $n_\up=1$ corresponds to a density of $1/3$ electron with spin $\up$ per atom.
\begin{align*}
    E(n_\up=0,n_\dn=0) &= 0 U + 0 V = 0 U' \\
    E(n_\up=1,n_\dn=0) &= 0 U + 3 \frac{1}{3}\frac{1}{3} V = 0 U' \\
    &= E(n_\up=0,n_\dn=1) \\    
    E(n_\up=1,n_\dn=1) &= 3 \frac{1}{3} \frac{1}{3} U + 3 \frac{2}{3}\frac{2}{3} V = U'
\end{align*}
Here, $U$ is the local Coulomb interaction in the atomic model, $V$ is the nearest-neighbor Coulomb interaction in the atomic model and $U'$ is the Hubbard interaction in the molecular model. Now, we want $E(n_\up=1,n_\dn=1)+E(n_\up=0,n_\dn=0)-E(n_\up=1,n_\dn=0)-E(n_\up=0,n_\dn=1)$ to be the same in both models, which requires $U' = \frac{1}{3} U + \frac{2}{3} V$. 
This geometric estimate gives $U'\approx2.1$ eV, while the full cRPA gives $U'\approx1.9$ eV, which means that the additional screening is responsible for roughly $0.2$ eV only. 
Note that more than half of the molecular Hubbard interaction comes from the interatomic Coulomb interaction $V$. This shows that a strictly Nb local Hubbard interaction is not a good approximation for the three-orbital atomic model, the interatomic interactions are just as important, similar to many two-dimensional hexagonal compounds~\cite{Schuler13}. 

For the inter-trimer interactions, we find slightly smaller values for the molecular model compared to the atomic model, which can also be understood from the fact that the molecular model contains additional screening processes. For longer ranges, these screening processes are less efficient, and the difference between the interactions in the two models is smaller.

\subsection{Hubbard-I}

A simple and lightweight way to take into account correlation effects on top of a density functional theory calculations is provided by the Hubbard-I approximation~\cite{Lichtenstein98}. For our compound, we first calculate the self-energy of a single, isolated trimer using all intratrimer Coulomb matrix elements. Then, we insert this local self-energy into the lattice Green's functions $G=G_0 /(1-\Sigma G_0)$, where $G_0$ is the Green's function corresponding to the DFT band structure. The resulting spectral functions $-\Im G(k,\omega)$ are shown throughout the manuscript. We use TRIQS~\cite{triqs} and the Hubbard-I solver implemented by Sch\"uler \textit{et al.}~\cite{schueler_hubbardi_2022} to perform the calculations.

There is always a certain amount of freedom in choosing the chemical potential when connecting band structures and many-electron calculations. For an insulator such as Nb$_3$Cl$_8$, this simply leads to an overall energy shift for all states/bands, which does not affect the physics as long as the chemical potential lies within the gap. To facilitate the comparison, we have lined out the spectra in different panels of several figures based on the center of the lowest band. 

\section{Data availability}
All data is available from the authors on reasonable request.

\section{Competing Interests}
The authors declare no competing financial or non-financial interests.

\section{Author Contributions}
S.G. and M.R. performed all ab initio calculations. E.v.L. analysed the trimer model. S.G., E.v.L. and M.R. performed the Hubbard-I calculations. M.R. designed the study with assistance from E.v.L. All authors contributed to the data analysis and the preparation of the manuscript.

\section{Acknowledgements}

We thank Mazhar Ali for drawing our attention to distorted kagome materials and for fruitful discussions and feedback on the manuscript.
E.v.L. acknowledges support from Gyllenstiernska Krapperupsstiftelsen, Crafoord Foundation and from the Swedish Research Council (Vetenskapsrådet, VR) under grant 2022-03090. M.R. and M.I.K. acknowledge support from the Dutch Research Council (NWO) via the “TOPCORE” consortium and from the research program “Materials for the Quantum Age” (QuMat) for financial support. The latter program (registration number 024.005.006) is part of the Gravitation program financed by the Dutch Ministry of Education, Culture and Science (OCW). The work of M.I.K. was further supported by the European Union’s Horizon 2020 research and innovation program under European Research Council Synergy Grant 854843 “FASTCORR”.

\bibliography{references}

%apsrev4-2.bst 2019-01-14 (MD) hand-edited version of apsrev4-1.bst
%Control: key (0)
%Control: author (8) initials jnrlst
%Control: editor formatted (1) identically to author
%Control: production of article title (0) allowed
%Control: page (0) single
%Control: year (1) truncated
%Control: production of eprint (0) enabled
\begin{thebibliography}{59}%
\makeatletter
\providecommand \@ifxundefined [1]{%
 \@ifx{#1\undefined}
}%
\providecommand \@ifnum [1]{%
 \ifnum #1\expandafter \@firstoftwo
 \else \expandafter \@secondoftwo
 \fi
}%
\providecommand \@ifx [1]{%
 \ifx #1\expandafter \@firstoftwo
 \else \expandafter \@secondoftwo
 \fi
}%
\providecommand \natexlab [1]{#1}%
\providecommand \enquote  [1]{``#1''}%
\providecommand \bibnamefont  [1]{#1}%
\providecommand \bibfnamefont [1]{#1}%
\providecommand \citenamefont [1]{#1}%
\providecommand \href@noop [0]{\@secondoftwo}%
\providecommand \href [0]{\begingroup \@sanitize@url \@href}%
\providecommand \@href[1]{\@@startlink{#1}\@@href}%
\providecommand \@@href[1]{\endgroup#1\@@endlink}%
\providecommand \@sanitize@url [0]{\catcode `\\12\catcode `\$12\catcode
  `\&12\catcode `\#12\catcode `\^12\catcode `\_12\catcode `\%12\relax}%
\providecommand \@@startlink[1]{}%
\providecommand \@@endlink[0]{}%
\providecommand \url  [0]{\begingroup\@sanitize@url \@url }%
\providecommand \@url [1]{\endgroup\@href {#1}{\urlprefix }}%
\providecommand \urlprefix  [0]{URL }%
\providecommand \Eprint [0]{\href }%
\providecommand \doibase [0]{https://doi.org/}%
\providecommand \selectlanguage [0]{\@gobble}%
\providecommand \bibinfo  [0]{\@secondoftwo}%
\providecommand \bibfield  [0]{\@secondoftwo}%
\providecommand \translation [1]{[#1]}%
\providecommand \BibitemOpen [0]{}%
\providecommand \bibitemStop [0]{}%
\providecommand \bibitemNoStop [0]{.\EOS\space}%
\providecommand \EOS [0]{\spacefactor3000\relax}%
\providecommand \BibitemShut  [1]{\csname bibitem#1\endcsname}%
\let\auto@bib@innerbib\@empty
%</preamble>
\bibitem [{\citenamefont {Schubin}\ and\ \citenamefont
  {Wonsowsky}(1934)}]{polar1934}%
  \BibitemOpen
  \bibfield  {author} {\bibinfo {author} {\bibfnamefont {S.}~\bibnamefont
  {Schubin}}\ and\ \bibinfo {author} {\bibfnamefont {S.}~\bibnamefont
  {Wonsowsky}},\ }\bibfield  {title} {\bibinfo {title} {On the electron theory
  of metals},\ }\href {https://doi.org/10.1098/rspa.1934.0089} {\bibfield
  {journal} {\bibinfo  {journal} {Proceedings of the Royal Society of London.
  Series A. Mathematical and Physical Sciences}\ }\textbf {\bibinfo {volume}
  {145}},\ \bibinfo {pages} {159} (\bibinfo {year} {1934})}\BibitemShut
  {NoStop}%
\bibitem [{\citenamefont {Vonsovsky}\ and\ \citenamefont
  {Katsnelson}(1979{\natexlab{a}})}]{VKI1979}%
  \BibitemOpen
  \bibfield  {author} {\bibinfo {author} {\bibfnamefont {S.~V.}\ \bibnamefont
  {Vonsovsky}}\ and\ \bibinfo {author} {\bibfnamefont {M.~I.}\ \bibnamefont
  {Katsnelson}},\ }\bibfield  {title} {\bibinfo {title} {Some types of
  instabilities in the electron energy spectrum of the polar model of the
  crystal. {I}. {The} maximum-polarity state},\ }\href
  {https://doi.org/10.1088/0022-3719/12/11/015} {\bibfield  {journal} {\bibinfo
   {journal} {Journal of Physics C: Solid State Physics}\ }\textbf {\bibinfo
  {volume} {12}},\ \bibinfo {pages} {2043} (\bibinfo {year}
  {1979}{\natexlab{a}})}\BibitemShut {NoStop}%
\bibitem [{\citenamefont {Vonsovsky}\ and\ \citenamefont
  {Katsnelson}(1979{\natexlab{b}})}]{VKII1979}%
  \BibitemOpen
  \bibfield  {author} {\bibinfo {author} {\bibfnamefont {S.~V.}\ \bibnamefont
  {Vonsovsky}}\ and\ \bibinfo {author} {\bibfnamefont {M.~I.}\ \bibnamefont
  {Katsnelson}},\ }\bibfield  {title} {\bibinfo {title} {Some types of
  instabilities in the electron energy spectrum of the polar model of the
  crystal. {II}. {The} criterion of stability of a metallic state},\ }\href
  {https://doi.org/10.1088/0022-3719/12/11/016} {\bibfield  {journal} {\bibinfo
   {journal} {Journal of Physics C: Solid State Physics}\ }\textbf {\bibinfo
  {volume} {12}},\ \bibinfo {pages} {2055} (\bibinfo {year}
  {1979}{\natexlab{b}})}\BibitemShut {NoStop}%
\bibitem [{\citenamefont {Gutzwiller}(1963)}]{Gutzwiller1963}%
  \BibitemOpen
  \bibfield  {author} {\bibinfo {author} {\bibfnamefont {M.~C.}\ \bibnamefont
  {Gutzwiller}},\ }\bibfield  {title} {\bibinfo {title} {Effect of correlation
  on the ferromagnetism of transition metals},\ }\href
  {https://doi.org/10.1103/PhysRevLett.10.159} {\bibfield  {journal} {\bibinfo
  {journal} {Phys. Rev. Lett.}\ }\textbf {\bibinfo {volume} {10}},\ \bibinfo
  {pages} {159} (\bibinfo {year} {1963})}\BibitemShut {NoStop}%
\bibitem [{\citenamefont {Kanamori}(1963)}]{Kanamori1963}%
  \BibitemOpen
  \bibfield  {author} {\bibinfo {author} {\bibfnamefont {J.}~\bibnamefont
  {Kanamori}},\ }\bibfield  {title} {\bibinfo {title} {{Electron Correlation
  and Ferromagnetism of Transition Metals}},\ }\href
  {https://doi.org/10.1143/PTP.30.275} {\bibfield  {journal} {\bibinfo
  {journal} {Progress of Theoretical Physics}\ }\textbf {\bibinfo {volume}
  {30}},\ \bibinfo {pages} {275} (\bibinfo {year} {1963})}\BibitemShut
  {NoStop}%
\bibitem [{\citenamefont {Hubbard}(1963)}]{Hubbard1963}%
  \BibitemOpen
  \bibfield  {author} {\bibinfo {author} {\bibfnamefont {J.}~\bibnamefont
  {Hubbard}},\ }\bibfield  {title} {\bibinfo {title} {Electron correlations in
  narrow energy bands},\ }\href {https://doi.org/10.1098/rspa.1963.0204}
  {\bibfield  {journal} {\bibinfo  {journal} {Proceedings of the Royal Society
  of London. Series A. Mathematical and Physical Sciences}\ }\textbf {\bibinfo
  {volume} {276}},\ \bibinfo {pages} {238} (\bibinfo {year}
  {1963})}\BibitemShut {NoStop}%
\bibitem [{\citenamefont {Hubbard}(1964{\natexlab{a}})}]{Hubbard1964}%
  \BibitemOpen
  \bibfield  {author} {\bibinfo {author} {\bibfnamefont {J.}~\bibnamefont
  {Hubbard}},\ }\bibfield  {title} {\bibinfo {title} {Electron correlations in
  narrow energy bands. {II}. {The} degenerate band case},\ }\href
  {https://doi.org/10.1098/rspa.1964.0019} {\bibfield  {journal} {\bibinfo
  {journal} {Proceedings of the Royal Society of London. Series A. Mathematical
  and Physical Sciences}\ }\textbf {\bibinfo {volume} {277}},\ \bibinfo {pages}
  {237} (\bibinfo {year} {1964}{\natexlab{a}})}\BibitemShut {NoStop}%
\bibitem [{\citenamefont {Hubbard}(1964{\natexlab{b}})}]{HubbardIII}%
  \BibitemOpen
  \bibfield  {author} {\bibinfo {author} {\bibfnamefont {J.}~\bibnamefont
  {Hubbard}},\ }\bibfield  {title} {\bibinfo {title} {Electron correlations in
  narrow energy bands. {III}. {An} improved solution},\ }\href
  {https://doi.org/10.1098/rspa.1964.0190} {\bibfield  {journal} {\bibinfo
  {journal} {Proceedings of the Royal Society of London. Series A. Mathematical
  and Physical Sciences}\ }\textbf {\bibinfo {volume} {281}},\ \bibinfo {pages}
  {401} (\bibinfo {year} {1964}{\natexlab{b}})}\BibitemShut {NoStop}%
\bibitem [{\citenamefont {Qin}\ \emph {et~al.}(2022)\citenamefont {Qin},
  \citenamefont {Sch\"{a}fer}, \citenamefont {Andergassen}, \citenamefont
  {Corboz},\ and\ \citenamefont {Gull}}]{Qin22}%
  \BibitemOpen
  \bibfield  {author} {\bibinfo {author} {\bibfnamefont {M.}~\bibnamefont
  {Qin}}, \bibinfo {author} {\bibfnamefont {T.}~\bibnamefont {Sch\"{a}fer}},
  \bibinfo {author} {\bibfnamefont {S.}~\bibnamefont {Andergassen}}, \bibinfo
  {author} {\bibfnamefont {P.}~\bibnamefont {Corboz}},\ and\ \bibinfo {author}
  {\bibfnamefont {E.}~\bibnamefont {Gull}},\ }\bibfield  {title} {\bibinfo
  {title} {The {Hubbard} model: A computational perspective},\ }\href
  {https://doi.org/10.1146/annurev-conmatphys-090921-033948} {\bibfield
  {journal} {\bibinfo  {journal} {Annual Review of Condensed Matter Physics}\
  }\textbf {\bibinfo {volume} {13}},\ \bibinfo {pages} {275} (\bibinfo {year}
  {2022})}\BibitemShut {NoStop}%
\bibitem [{\citenamefont {Lieb}\ and\ \citenamefont {Wu}(1968)}]{Lieb1968}%
  \BibitemOpen
  \bibfield  {author} {\bibinfo {author} {\bibfnamefont {E.~H.}\ \bibnamefont
  {Lieb}}\ and\ \bibinfo {author} {\bibfnamefont {F.~Y.}\ \bibnamefont {Wu}},\
  }\bibfield  {title} {\bibinfo {title} {Absence of {Mott} transition in an
  exact solution of the short-range, one-band model in one dimension},\ }\href
  {https://doi.org/10.1103/PhysRevLett.20.1445} {\bibfield  {journal} {\bibinfo
   {journal} {Phys. Rev. Lett.}\ }\textbf {\bibinfo {volume} {20}},\ \bibinfo
  {pages} {1445} (\bibinfo {year} {1968})}\BibitemShut {NoStop}%
\bibitem [{\citenamefont {Metzner}\ and\ \citenamefont
  {Vollhardt}(1989)}]{Metzner89}%
  \BibitemOpen
  \bibfield  {author} {\bibinfo {author} {\bibfnamefont {W.}~\bibnamefont
  {Metzner}}\ and\ \bibinfo {author} {\bibfnamefont {D.}~\bibnamefont
  {Vollhardt}},\ }\bibfield  {title} {\bibinfo {title} {Correlated lattice
  fermions in $d=\ensuremath{\infty}$ dimensions},\ }\href
  {https://doi.org/10.1103/PhysRevLett.62.324} {\bibfield  {journal} {\bibinfo
  {journal} {Phys. Rev. Lett.}\ }\textbf {\bibinfo {volume} {62}},\ \bibinfo
  {pages} {324} (\bibinfo {year} {1989})}\BibitemShut {NoStop}%
\bibitem [{\citenamefont {Georges}\ and\ \citenamefont
  {Kotliar}(1992)}]{Georges92}%
  \BibitemOpen
  \bibfield  {author} {\bibinfo {author} {\bibfnamefont {A.}~\bibnamefont
  {Georges}}\ and\ \bibinfo {author} {\bibfnamefont {G.}~\bibnamefont
  {Kotliar}},\ }\bibfield  {title} {\bibinfo {title} {Hubbard model in infinite
  dimensions},\ }\href {https://doi.org/10.1103/PhysRevB.45.6479} {\bibfield
  {journal} {\bibinfo  {journal} {Phys. Rev. B}\ }\textbf {\bibinfo {volume}
  {45}},\ \bibinfo {pages} {6479} (\bibinfo {year} {1992})}\BibitemShut
  {NoStop}%
\bibitem [{\citenamefont {LeBlanc}\ \emph {et~al.}(2015)\citenamefont
  {LeBlanc}, \citenamefont {Antipov}, \citenamefont {Becca}, \citenamefont
  {Bulik}, \citenamefont {Chan}, \citenamefont {Chung}, \citenamefont {Deng},
  \citenamefont {Ferrero}, \citenamefont {Henderson}, \citenamefont
  {Jim\'enez-Hoyos}, \citenamefont {Kozik}, \citenamefont {Liu}, \citenamefont
  {Millis}, \citenamefont {Prokof'ev}, \citenamefont {Qin}, \citenamefont
  {Scuseria}, \citenamefont {Shi}, \citenamefont {Svistunov}, \citenamefont
  {Tocchio}, \citenamefont {Tupitsyn}, \citenamefont {White}, \citenamefont
  {Zhang}, \citenamefont {Zheng}, \citenamefont {Zhu},\ and\ \citenamefont
  {Gull}}]{LeBlanc15}%
  \BibitemOpen
  \bibfield  {author} {\bibinfo {author} {\bibfnamefont {J.~P.~F.}\
  \bibnamefont {LeBlanc}}, \bibinfo {author} {\bibfnamefont {A.~E.}\
  \bibnamefont {Antipov}}, \bibinfo {author} {\bibfnamefont {F.}~\bibnamefont
  {Becca}}, \bibinfo {author} {\bibfnamefont {I.~W.}\ \bibnamefont {Bulik}},
  \bibinfo {author} {\bibfnamefont {G.~K.-L.}\ \bibnamefont {Chan}}, \bibinfo
  {author} {\bibfnamefont {C.-M.}\ \bibnamefont {Chung}}, \bibinfo {author}
  {\bibfnamefont {Y.}~\bibnamefont {Deng}}, \bibinfo {author} {\bibfnamefont
  {M.}~\bibnamefont {Ferrero}}, \bibinfo {author} {\bibfnamefont {T.~M.}\
  \bibnamefont {Henderson}}, \bibinfo {author} {\bibfnamefont {C.~A.}\
  \bibnamefont {Jim\'enez-Hoyos}}, \bibinfo {author} {\bibfnamefont
  {E.}~\bibnamefont {Kozik}}, \bibinfo {author} {\bibfnamefont {X.-W.}\
  \bibnamefont {Liu}}, \bibinfo {author} {\bibfnamefont {A.~J.}\ \bibnamefont
  {Millis}}, \bibinfo {author} {\bibfnamefont {N.~V.}\ \bibnamefont
  {Prokof'ev}}, \bibinfo {author} {\bibfnamefont {M.}~\bibnamefont {Qin}},
  \bibinfo {author} {\bibfnamefont {G.~E.}\ \bibnamefont {Scuseria}}, \bibinfo
  {author} {\bibfnamefont {H.}~\bibnamefont {Shi}}, \bibinfo {author}
  {\bibfnamefont {B.~V.}\ \bibnamefont {Svistunov}}, \bibinfo {author}
  {\bibfnamefont {L.~F.}\ \bibnamefont {Tocchio}}, \bibinfo {author}
  {\bibfnamefont {I.~S.}\ \bibnamefont {Tupitsyn}}, \bibinfo {author}
  {\bibfnamefont {S.~R.}\ \bibnamefont {White}}, \bibinfo {author}
  {\bibfnamefont {S.}~\bibnamefont {Zhang}}, \bibinfo {author} {\bibfnamefont
  {B.-X.}\ \bibnamefont {Zheng}}, \bibinfo {author} {\bibfnamefont
  {Z.}~\bibnamefont {Zhu}},\ and\ \bibinfo {author} {\bibfnamefont
  {E.}~\bibnamefont {Gull}} (\bibinfo {collaboration} {Simons Collaboration on
  the Many-Electron Problem}),\ }\bibfield  {title} {\bibinfo {title}
  {Solutions of the two-dimensional {Hubbard} model: Benchmarks and results
  from a wide range of numerical algorithms},\ }\href
  {https://doi.org/10.1103/PhysRevX.5.041041} {\bibfield  {journal} {\bibinfo
  {journal} {Phys. Rev. X}\ }\textbf {\bibinfo {volume} {5}},\ \bibinfo {pages}
  {041041} (\bibinfo {year} {2015})}\BibitemShut {NoStop}%
\bibitem [{\citenamefont {Arovas}\ \emph {et~al.}(2022)\citenamefont {Arovas},
  \citenamefont {Berg}, \citenamefont {Kivelson},\ and\ \citenamefont
  {Raghu}}]{Arovas22}%
  \BibitemOpen
  \bibfield  {author} {\bibinfo {author} {\bibfnamefont {D.~P.}\ \bibnamefont
  {Arovas}}, \bibinfo {author} {\bibfnamefont {E.}~\bibnamefont {Berg}},
  \bibinfo {author} {\bibfnamefont {S.~A.}\ \bibnamefont {Kivelson}},\ and\
  \bibinfo {author} {\bibfnamefont {S.}~\bibnamefont {Raghu}},\ }\bibfield
  {title} {\bibinfo {title} {The hubbard model},\ }\href
  {https://doi.org/10.1146/annurev-conmatphys-031620-102024} {\bibfield
  {journal} {\bibinfo  {journal} {Annual Review of Condensed Matter Physics}\
  }\textbf {\bibinfo {volume} {13}},\ \bibinfo {pages} {239} (\bibinfo {year}
  {2022})}\BibitemShut {NoStop}%
\bibitem [{\citenamefont {Andersen}\ \emph {et~al.}(1995)\citenamefont
  {Andersen}, \citenamefont {Liechtenstein}, \citenamefont {Jepsen},\ and\
  \citenamefont {Paulsen}}]{OKA}%
  \BibitemOpen
  \bibfield  {author} {\bibinfo {author} {\bibfnamefont {O.~K.}\ \bibnamefont
  {Andersen}}, \bibinfo {author} {\bibfnamefont {A.~I.}\ \bibnamefont
  {Liechtenstein}}, \bibinfo {author} {\bibfnamefont {O.}~\bibnamefont
  {Jepsen}},\ and\ \bibinfo {author} {\bibfnamefont {F.}~\bibnamefont
  {Paulsen}},\ }\bibfield  {title} {\bibinfo {title} {{L}{D}{A} energy bands,
  low-energy hamiltonians, $t^\prime$, $t^{\prime\prime}$, t$_{\perp}$(k), and
  {J}$_{\perp}$},\ }\href {https://doi.org/10.1016/0022-3697(95)00269-3}
  {\bibfield  {journal} {\bibinfo  {journal} {J. Phys. Chem. Solids}\ }\textbf
  {\bibinfo {volume} {56}},\ \bibinfo {pages} {1573} (\bibinfo {year}
  {1995})}\BibitemShut {NoStop}%
\bibitem [{\citenamefont {Pavarini}\ \emph {et~al.}(2001)\citenamefont
  {Pavarini}, \citenamefont {Dasgupta}, \citenamefont {Saha-Dasgupta},
  \citenamefont {Jepsen},\ and\ \citenamefont {Andersen}}]{Pavarini}%
  \BibitemOpen
  \bibfield  {author} {\bibinfo {author} {\bibfnamefont {E.}~\bibnamefont
  {Pavarini}}, \bibinfo {author} {\bibfnamefont {I.}~\bibnamefont {Dasgupta}},
  \bibinfo {author} {\bibfnamefont {T.}~\bibnamefont {Saha-Dasgupta}}, \bibinfo
  {author} {\bibfnamefont {O.}~\bibnamefont {Jepsen}},\ and\ \bibinfo {author}
  {\bibfnamefont {O.~K.}\ \bibnamefont {Andersen}},\ }\bibfield  {title}
  {\bibinfo {title} {Band-structure trend in hole-doped cuprates and
  correlation with {$T_c^{\rm max}$}},\ }\href
  {https://doi.org/10.1103/PhysRevLett.87.047003} {\bibfield  {journal}
  {\bibinfo  {journal} {Phys. Rev. Lett.}\ }\textbf {\bibinfo {volume} {87}},\
  \bibinfo {pages} {047003} (\bibinfo {year} {2001})}\BibitemShut {NoStop}%
\bibitem [{\citenamefont {Lechermann}(2020)}]{Lechermann20}%
  \BibitemOpen
  \bibfield  {author} {\bibinfo {author} {\bibfnamefont {F.}~\bibnamefont
  {Lechermann}},\ }\bibfield  {title} {\bibinfo {title} {Late transition metal
  oxides with infinite-layer structure: Nickelates versus cuprates},\ }\href
  {https://doi.org/10.1103/PhysRevB.101.081110} {\bibfield  {journal} {\bibinfo
   {journal} {Phys. Rev. B}\ }\textbf {\bibinfo {volume} {101}},\ \bibinfo
  {pages} {081110} (\bibinfo {year} {2020})}\BibitemShut {NoStop}%
\bibitem [{\citenamefont {Kitatani}\ \emph {et~al.}(2020)\citenamefont
  {Kitatani}, \citenamefont {Si}, \citenamefont {Janson}, \citenamefont
  {Arita}, \citenamefont {Zhong},\ and\ \citenamefont {Held}}]{Kitatani20}%
  \BibitemOpen
  \bibfield  {author} {\bibinfo {author} {\bibfnamefont {M.}~\bibnamefont
  {Kitatani}}, \bibinfo {author} {\bibfnamefont {L.}~\bibnamefont {Si}},
  \bibinfo {author} {\bibfnamefont {O.}~\bibnamefont {Janson}}, \bibinfo
  {author} {\bibfnamefont {R.}~\bibnamefont {Arita}}, \bibinfo {author}
  {\bibfnamefont {Z.}~\bibnamefont {Zhong}},\ and\ \bibinfo {author}
  {\bibfnamefont {K.}~\bibnamefont {Held}},\ }\bibfield  {title} {\bibinfo
  {title} {Nickelate superconductors—a renaissance of the one-band hubbard
  model},\ }\href {https://doi.org/10.1038/s41535-020-00260-y} {\bibfield
  {journal} {\bibinfo  {journal} {npj Quantum Materials}\ }\textbf {\bibinfo
  {volume} {5}},\ \bibinfo {pages} {59} (\bibinfo {year} {2020})}\BibitemShut
  {NoStop}%
\bibitem [{\citenamefont {Wilhelm}\ \emph {et~al.}(2015)\citenamefont
  {Wilhelm}, \citenamefont {Lechermann}, \citenamefont {Hafermann},
  \citenamefont {Katsnelson},\ and\ \citenamefont
  {Lichtenstein}}]{Lechermann2015}%
  \BibitemOpen
  \bibfield  {author} {\bibinfo {author} {\bibfnamefont {A.}~\bibnamefont
  {Wilhelm}}, \bibinfo {author} {\bibfnamefont {F.}~\bibnamefont {Lechermann}},
  \bibinfo {author} {\bibfnamefont {H.}~\bibnamefont {Hafermann}}, \bibinfo
  {author} {\bibfnamefont {M.~I.}\ \bibnamefont {Katsnelson}},\ and\ \bibinfo
  {author} {\bibfnamefont {A.~I.}\ \bibnamefont {Lichtenstein}},\ }\bibfield
  {title} {\bibinfo {title} {From {Hubbard} bands to spin-polaron excitations
  in the doped {Mott} material {${\mathrm{Na}}_{x}{\mathrm{CoO}}_{2}$}},\
  }\href {https://doi.org/10.1103/PhysRevB.91.155114} {\bibfield  {journal}
  {\bibinfo  {journal} {Phys. Rev. B}\ }\textbf {\bibinfo {volume} {91}},\
  \bibinfo {pages} {155114} (\bibinfo {year} {2015})}\BibitemShut {NoStop}%
\bibitem [{\citenamefont {Perfetti}\ \emph {et~al.}(2003)\citenamefont
  {Perfetti}, \citenamefont {Georges}, \citenamefont {Florens}, \citenamefont
  {Biermann}, \citenamefont {Mitrovic}, \citenamefont {Berger}, \citenamefont
  {Tomm}, \citenamefont {H\"ochst},\ and\ \citenamefont {Grioni}}]{Perfetti03}%
  \BibitemOpen
  \bibfield  {author} {\bibinfo {author} {\bibfnamefont {L.}~\bibnamefont
  {Perfetti}}, \bibinfo {author} {\bibfnamefont {A.}~\bibnamefont {Georges}},
  \bibinfo {author} {\bibfnamefont {S.}~\bibnamefont {Florens}}, \bibinfo
  {author} {\bibfnamefont {S.}~\bibnamefont {Biermann}}, \bibinfo {author}
  {\bibfnamefont {S.}~\bibnamefont {Mitrovic}}, \bibinfo {author}
  {\bibfnamefont {H.}~\bibnamefont {Berger}}, \bibinfo {author} {\bibfnamefont
  {Y.}~\bibnamefont {Tomm}}, \bibinfo {author} {\bibfnamefont {H.}~\bibnamefont
  {H\"ochst}},\ and\ \bibinfo {author} {\bibfnamefont {M.}~\bibnamefont
  {Grioni}},\ }\bibfield  {title} {\bibinfo {title} {Spectroscopic signatures
  of a bandwidth-controlled mott transition at the surface of
  $1t\mathrm{\text{\ensuremath{-}}}{\mathrm{t}\mathrm{a}\mathrm{s}\mathrm{e}}_{2}$},\
  }\href {https://doi.org/10.1103/PhysRevLett.90.166401} {\bibfield  {journal}
  {\bibinfo  {journal} {Phys. Rev. Lett.}\ }\textbf {\bibinfo {volume} {90}},\
  \bibinfo {pages} {166401} (\bibinfo {year} {2003})}\BibitemShut {NoStop}%
\bibitem [{\citenamefont {Hansmann}\ \emph
  {et~al.}(2013{\natexlab{a}})\citenamefont {Hansmann}, \citenamefont {Ayral},
  \citenamefont {Vaugier}, \citenamefont {Werner},\ and\ \citenamefont
  {Biermann}}]{Hansmann13}%
  \BibitemOpen
  \bibfield  {author} {\bibinfo {author} {\bibfnamefont {P.}~\bibnamefont
  {Hansmann}}, \bibinfo {author} {\bibfnamefont {T.}~\bibnamefont {Ayral}},
  \bibinfo {author} {\bibfnamefont {L.}~\bibnamefont {Vaugier}}, \bibinfo
  {author} {\bibfnamefont {P.}~\bibnamefont {Werner}},\ and\ \bibinfo {author}
  {\bibfnamefont {S.}~\bibnamefont {Biermann}},\ }\bibfield  {title} {\bibinfo
  {title} {Long-range coulomb interactions in surface systems: A
  first-principles description within self-consistently combined $gw$ and
  dynamical mean-field theory},\ }\href
  {https://doi.org/10.1103/PhysRevLett.110.166401} {\bibfield  {journal}
  {\bibinfo  {journal} {Phys. Rev. Lett.}\ }\textbf {\bibinfo {volume} {110}},\
  \bibinfo {pages} {166401} (\bibinfo {year} {2013}{\natexlab{a}})}\BibitemShut
  {NoStop}%
\bibitem [{\citenamefont {Hansmann}\ \emph
  {et~al.}(2013{\natexlab{b}})\citenamefont {Hansmann}, \citenamefont
  {Vaugier}, \citenamefont {Jiang},\ and\ \citenamefont
  {Biermann}}]{Hansmann13b}%
  \BibitemOpen
  \bibfield  {author} {\bibinfo {author} {\bibfnamefont {P.}~\bibnamefont
  {Hansmann}}, \bibinfo {author} {\bibfnamefont {L.}~\bibnamefont {Vaugier}},
  \bibinfo {author} {\bibfnamefont {H.}~\bibnamefont {Jiang}},\ and\ \bibinfo
  {author} {\bibfnamefont {S.}~\bibnamefont {Biermann}},\ }\bibfield  {title}
  {\bibinfo {title} {What about u on surfaces? {Extended} {Hubbard} models for
  adatom systems from first principles},\ }\href
  {https://doi.org/10.1088/0953-8984/25/9/094005} {\bibfield  {journal}
  {\bibinfo  {journal} {Journal of Physics: Condensed Matter}\ }\textbf
  {\bibinfo {volume} {25}},\ \bibinfo {pages} {094005} (\bibinfo {year}
  {2013}{\natexlab{b}})}\BibitemShut {NoStop}%
\bibitem [{\citenamefont {Esslinger}(2010)}]{Esslinger10}%
  \BibitemOpen
  \bibfield  {author} {\bibinfo {author} {\bibfnamefont {T.}~\bibnamefont
  {Esslinger}},\ }\bibfield  {title} {\bibinfo {title} {Fermi-hubbard physics
  with atoms in an optical lattice},\ }\href
  {https://doi.org/10.1146/annurev-conmatphys-070909-104059} {\bibfield
  {journal} {\bibinfo  {journal} {Annual Review of Condensed Matter Physics}\
  }\textbf {\bibinfo {volume} {1}},\ \bibinfo {pages} {129} (\bibinfo {year}
  {2010})}\BibitemShut {NoStop}%
\bibitem [{\citenamefont {Sch\"uler}\ \emph {et~al.}(2013)\citenamefont
  {Sch\"uler}, \citenamefont {R\"osner}, \citenamefont {Wehling}, \citenamefont
  {Lichtenstein},\ and\ \citenamefont {Katsnelson}}]{Schuler13}%
  \BibitemOpen
  \bibfield  {author} {\bibinfo {author} {\bibfnamefont {M.}~\bibnamefont
  {Sch\"uler}}, \bibinfo {author} {\bibfnamefont {M.}~\bibnamefont {R\"osner}},
  \bibinfo {author} {\bibfnamefont {T.~O.}\ \bibnamefont {Wehling}}, \bibinfo
  {author} {\bibfnamefont {A.~I.}\ \bibnamefont {Lichtenstein}},\ and\ \bibinfo
  {author} {\bibfnamefont {M.~I.}\ \bibnamefont {Katsnelson}},\ }\bibfield
  {title} {\bibinfo {title} {Optimal hubbard models for materials with nonlocal
  coulomb interactions: Graphene, silicene, and benzene},\ }\href
  {https://doi.org/10.1103/PhysRevLett.111.036601} {\bibfield  {journal}
  {\bibinfo  {journal} {Phys. Rev. Lett.}\ }\textbf {\bibinfo {volume} {111}},\
  \bibinfo {pages} {036601} (\bibinfo {year} {2013})}\BibitemShut {NoStop}%
\bibitem [{\citenamefont {Wang}\ \emph {et~al.}(2023)\citenamefont {Wang},
  \citenamefont {Wu}, \citenamefont {{McCandless}}, \citenamefont {Chan},\ and\
  \citenamefont {Ali}}]{wang_quantum_2023}%
  \BibitemOpen
  \bibfield  {author} {\bibinfo {author} {\bibfnamefont {Y.}~\bibnamefont
  {Wang}}, \bibinfo {author} {\bibfnamefont {H.}~\bibnamefont {Wu}}, \bibinfo
  {author} {\bibfnamefont {G.~T.}\ \bibnamefont {{McCandless}}}, \bibinfo
  {author} {\bibfnamefont {J.~Y.}\ \bibnamefont {Chan}},\ and\ \bibinfo
  {author} {\bibfnamefont {M.~N.}\ \bibnamefont {Ali}},\ }\href
  {https://doi.org/10.48550/arXiv.2303.03359} {\bibinfo {title} {Quantum states
  and intertwining phases in kagome materials}} (\bibinfo {year} {2023}),\
  \Eprint {https://arxiv.org/abs/2303.03359} {arXiv:2303.03359} \BibitemShut
  {NoStop}%
\bibitem [{\citenamefont {Gao}\ \emph {et~al.}(2022)\citenamefont {Gao},
  \citenamefont {Zhang}, \citenamefont {Wang}, \citenamefont {Tao},
  \citenamefont {Liu}, \citenamefont {Wang}, \citenamefont {Yuan},
  \citenamefont {Qu}, \citenamefont {Pan}, \citenamefont {Peng}, \citenamefont
  {Hu}, \citenamefont {Li}, \citenamefont {Huang}, \citenamefont {Zhou},
  \citenamefont {Meng}, \citenamefont {Yang}, \citenamefont {Wang},
  \citenamefont {Yao}, \citenamefont {Chen}, \citenamefont {Shi}, \citenamefont
  {Ding}, \citenamefont {Jiang}, \citenamefont {Li}, \citenamefont {Shi},
  \citenamefont {Weng},\ and\ \citenamefont {Qian}}]{Gao2022}%
  \BibitemOpen
  \bibfield  {author} {\bibinfo {author} {\bibfnamefont {S.}~\bibnamefont
  {Gao}}, \bibinfo {author} {\bibfnamefont {S.}~\bibnamefont {Zhang}}, \bibinfo
  {author} {\bibfnamefont {C.}~\bibnamefont {Wang}}, \bibinfo {author}
  {\bibfnamefont {W.}~\bibnamefont {Tao}}, \bibinfo {author} {\bibfnamefont
  {J.}~\bibnamefont {Liu}}, \bibinfo {author} {\bibfnamefont {T.}~\bibnamefont
  {Wang}}, \bibinfo {author} {\bibfnamefont {S.}~\bibnamefont {Yuan}}, \bibinfo
  {author} {\bibfnamefont {G.}~\bibnamefont {Qu}}, \bibinfo {author}
  {\bibfnamefont {M.}~\bibnamefont {Pan}}, \bibinfo {author} {\bibfnamefont
  {S.}~\bibnamefont {Peng}}, \bibinfo {author} {\bibfnamefont {Y.}~\bibnamefont
  {Hu}}, \bibinfo {author} {\bibfnamefont {H.}~\bibnamefont {Li}}, \bibinfo
  {author} {\bibfnamefont {Y.}~\bibnamefont {Huang}}, \bibinfo {author}
  {\bibfnamefont {H.}~\bibnamefont {Zhou}}, \bibinfo {author} {\bibfnamefont
  {S.}~\bibnamefont {Meng}}, \bibinfo {author} {\bibfnamefont {L.}~\bibnamefont
  {Yang}}, \bibinfo {author} {\bibfnamefont {Z.}~\bibnamefont {Wang}}, \bibinfo
  {author} {\bibfnamefont {Y.}~\bibnamefont {Yao}}, \bibinfo {author}
  {\bibfnamefont {Z.}~\bibnamefont {Chen}}, \bibinfo {author} {\bibfnamefont
  {M.}~\bibnamefont {Shi}}, \bibinfo {author} {\bibfnamefont {H.}~\bibnamefont
  {Ding}}, \bibinfo {author} {\bibfnamefont {K.}~\bibnamefont {Jiang}},
  \bibinfo {author} {\bibfnamefont {Y.}~\bibnamefont {Li}}, \bibinfo {author}
  {\bibfnamefont {Y.}~\bibnamefont {Shi}}, \bibinfo {author} {\bibfnamefont
  {H.}~\bibnamefont {Weng}},\ and\ \bibinfo {author} {\bibfnamefont
  {T.}~\bibnamefont {Qian}},\ }\href@noop {} {\bibinfo {title} {Mott insulator
  state in a van der waals flat-band compound}} (\bibinfo {year} {2022}),\
  \Eprint {https://arxiv.org/abs/2205.11462} {arXiv:2205.11462
  [cond-mat.str-el]} \BibitemShut {NoStop}%
\bibitem [{\citenamefont {Zhang}\ \emph {et~al.}(2023)\citenamefont {Zhang},
  \citenamefont {Gu}, \citenamefont {Weng}, \citenamefont {Jiang},\ and\
  \citenamefont {Hu}}]{Zhang23}%
  \BibitemOpen
  \bibfield  {author} {\bibinfo {author} {\bibfnamefont {Y.}~\bibnamefont
  {Zhang}}, \bibinfo {author} {\bibfnamefont {Y.}~\bibnamefont {Gu}}, \bibinfo
  {author} {\bibfnamefont {H.}~\bibnamefont {Weng}}, \bibinfo {author}
  {\bibfnamefont {K.}~\bibnamefont {Jiang}},\ and\ \bibinfo {author}
  {\bibfnamefont {J.}~\bibnamefont {Hu}},\ }\bibfield  {title} {\bibinfo
  {title} {Mottness in two-dimensional van der waals ${\mathrm{nb}}_{3}{X}_{8}$
  monolayers
  $(x=\mathrm{Cl},\mathrm{Br},\phantom{\rule{0.28em}{0ex}}\text{and}\phantom{\rule{0.28em}{0ex}}\mathrm{I})$},\
  }\href {https://doi.org/10.1103/PhysRevB.107.035126} {\bibfield  {journal}
  {\bibinfo  {journal} {Phys. Rev. B}\ }\textbf {\bibinfo {volume} {107}},\
  \bibinfo {pages} {035126} (\bibinfo {year} {2023})}\BibitemShut {NoStop}%
\bibitem [{\citenamefont {Sun}\ \emph {et~al.}(2022)\citenamefont {Sun},
  \citenamefont {Zhou}, \citenamefont {Wang}, \citenamefont {Kumar},
  \citenamefont {Geng}, \citenamefont {Yue}, \citenamefont {Han}, \citenamefont
  {Haraguchi}, \citenamefont {Shimada}, \citenamefont {Cheng}, \citenamefont
  {Chen}, \citenamefont {Shi}, \citenamefont {Wu}, \citenamefont {Meng},\ and\
  \citenamefont {Feng}}]{sun_observation_2022}%
  \BibitemOpen
  \bibfield  {author} {\bibinfo {author} {\bibfnamefont {Z.}~\bibnamefont
  {Sun}}, \bibinfo {author} {\bibfnamefont {H.}~\bibnamefont {Zhou}}, \bibinfo
  {author} {\bibfnamefont {C.}~\bibnamefont {Wang}}, \bibinfo {author}
  {\bibfnamefont {S.}~\bibnamefont {Kumar}}, \bibinfo {author} {\bibfnamefont
  {D.}~\bibnamefont {Geng}}, \bibinfo {author} {\bibfnamefont {S.}~\bibnamefont
  {Yue}}, \bibinfo {author} {\bibfnamefont {X.}~\bibnamefont {Han}}, \bibinfo
  {author} {\bibfnamefont {Y.}~\bibnamefont {Haraguchi}}, \bibinfo {author}
  {\bibfnamefont {K.}~\bibnamefont {Shimada}}, \bibinfo {author} {\bibfnamefont
  {P.}~\bibnamefont {Cheng}}, \bibinfo {author} {\bibfnamefont
  {L.}~\bibnamefont {Chen}}, \bibinfo {author} {\bibfnamefont {Y.}~\bibnamefont
  {Shi}}, \bibinfo {author} {\bibfnamefont {K.}~\bibnamefont {Wu}}, \bibinfo
  {author} {\bibfnamefont {S.}~\bibnamefont {Meng}},\ and\ \bibinfo {author}
  {\bibfnamefont {B.}~\bibnamefont {Feng}},\ }\bibfield  {title} {\bibinfo
  {title} {Observation of topological flat bands in the kagome semiconductor
  nb3cl8},\ }\href {https://doi.org/10.1021/acs.nanolett.2c00778} {\bibfield
  {journal} {\bibinfo  {journal} {Nano Letters}\ }\textbf {\bibinfo {volume}
  {22}},\ \bibinfo {pages} {4596} (\bibinfo {year} {2022})}\BibitemShut
  {NoStop}%
\bibitem [{\citenamefont {Yoon}\ \emph {et~al.}(2020)\citenamefont {Yoon},
  \citenamefont {Lesne}, \citenamefont {Sklarek}, \citenamefont {Sheckelton},
  \citenamefont {Pasco}, \citenamefont {Parkin}, \citenamefont {McQueen},\ and\
  \citenamefont {Ali}}]{Yoon2020}%
  \BibitemOpen
  \bibfield  {author} {\bibinfo {author} {\bibfnamefont {J.}~\bibnamefont
  {Yoon}}, \bibinfo {author} {\bibfnamefont {E.}~\bibnamefont {Lesne}},
  \bibinfo {author} {\bibfnamefont {K.}~\bibnamefont {Sklarek}}, \bibinfo
  {author} {\bibfnamefont {J.}~\bibnamefont {Sheckelton}}, \bibinfo {author}
  {\bibfnamefont {C.}~\bibnamefont {Pasco}}, \bibinfo {author} {\bibfnamefont
  {S.~S.~P.}\ \bibnamefont {Parkin}}, \bibinfo {author} {\bibfnamefont {T.~M.}\
  \bibnamefont {McQueen}},\ and\ \bibinfo {author} {\bibfnamefont {M.~N.}\
  \bibnamefont {Ali}},\ }\bibfield  {title} {\bibinfo {title} {Anomalous
  thickness-dependent electrical conductivity in van der waals layered
  transition metal halide, nb3cl8},\ }\href
  {https://doi.org/10.1088/1361-648X/ab832b} {\bibfield  {journal} {\bibinfo
  {journal} {Journal of Physics: Condensed Matter}\ }\textbf {\bibinfo {volume}
  {32}},\ \bibinfo {pages} {304004} (\bibinfo {year} {2020})}\BibitemShut
  {NoStop}%
\bibitem [{\citenamefont {Aryasetiawan}\ \emph {et~al.}(2004)\citenamefont
  {Aryasetiawan}, \citenamefont {Imada}, \citenamefont {Georges}, \citenamefont
  {Kotliar}, \citenamefont {Biermann},\ and\ \citenamefont
  {Lichtenstein}}]{Aryasetiawan04}%
  \BibitemOpen
  \bibfield  {author} {\bibinfo {author} {\bibfnamefont {F.}~\bibnamefont
  {Aryasetiawan}}, \bibinfo {author} {\bibfnamefont {M.}~\bibnamefont {Imada}},
  \bibinfo {author} {\bibfnamefont {A.}~\bibnamefont {Georges}}, \bibinfo
  {author} {\bibfnamefont {G.}~\bibnamefont {Kotliar}}, \bibinfo {author}
  {\bibfnamefont {S.}~\bibnamefont {Biermann}},\ and\ \bibinfo {author}
  {\bibfnamefont {A.~I.}\ \bibnamefont {Lichtenstein}},\ }\bibfield  {title}
  {\bibinfo {title} {Frequency-dependent local interactions and low-energy
  effective models from electronic structure calculations},\ }\href
  {https://doi.org/10.1103/PhysRevB.70.195104} {\bibfield  {journal} {\bibinfo
  {journal} {Phys. Rev. B}\ }\textbf {\bibinfo {volume} {70}},\ \bibinfo
  {pages} {195104} (\bibinfo {year} {2004})}\BibitemShut {NoStop}%
\bibitem [{\citenamefont {{Sheckelton John
  P.}}(2017)}]{sheckelton_john_p_rearrangement_2017}%
  \BibitemOpen
  \bibfield  {author} {\bibinfo {author} {\bibnamefont {{Sheckelton John
  P.}}},\ }\bibfield  {title} {\bibinfo {title} {Rearrangement of van der waals
  stacking and formation of a singlet state at t = 90 k in a cluster magnet},\
  }\href {https://doi.org/https://doi.org/10.1039/C6QI00470A} {\bibfield
  {journal} {\bibinfo  {journal} {Inorg. Chem. Front.}\ }\textbf {\bibinfo
  {volume} {4}},\ \bibinfo {pages} {481} (\bibinfo {year} {2017})}\BibitemShut
  {NoStop}%
\bibitem [{\citenamefont {Haraguchi}\ \emph {et~al.}(2017)\citenamefont
  {Haraguchi}, \citenamefont {Michioka}, \citenamefont {Ishikawa},
  \citenamefont {Nakano}, \citenamefont {Yamochi}, \citenamefont {Ueda},\ and\
  \citenamefont {Yoshimura}}]{haraguchi_magneticnonmagnetic_2017}%
  \BibitemOpen
  \bibfield  {author} {\bibinfo {author} {\bibfnamefont {Y.}~\bibnamefont
  {Haraguchi}}, \bibinfo {author} {\bibfnamefont {C.}~\bibnamefont {Michioka}},
  \bibinfo {author} {\bibfnamefont {M.}~\bibnamefont {Ishikawa}}, \bibinfo
  {author} {\bibfnamefont {Y.}~\bibnamefont {Nakano}}, \bibinfo {author}
  {\bibfnamefont {H.}~\bibnamefont {Yamochi}}, \bibinfo {author} {\bibfnamefont
  {H.}~\bibnamefont {Ueda}},\ and\ \bibinfo {author} {\bibfnamefont
  {K.}~\bibnamefont {Yoshimura}},\ }\bibfield  {title} {\bibinfo {title}
  {{Magnetic–Nonmagnetic Phase Transition with Interlayer Charge
  Disproportionation of Nb3 Trimers in the Cluster Compound Nb3Cl8}},\ }\href
  {https://doi.org/10.1021/acs.inorgchem.6b03028} {\bibfield  {journal}
  {\bibinfo  {journal} {Inorganic Chemistry}\ }\textbf {\bibinfo {volume}
  {56}},\ \bibinfo {pages} {3483} (\bibinfo {year} {2017})}\BibitemShut
  {NoStop}%
\bibitem [{\citenamefont {Kim}\ \emph {et~al.}(2023)\citenamefont {Kim},
  \citenamefont {Lee}, \citenamefont {Choi}, \citenamefont {Jung},
  \citenamefont {Son}, \citenamefont {Kim}, \citenamefont {Choi}, \citenamefont
  {Park},\ and\ \citenamefont {Kim}}]{kim_terahertz_2023}%
  \BibitemOpen
  \bibfield  {author} {\bibinfo {author} {\bibfnamefont {J.}~\bibnamefont
  {Kim}}, \bibinfo {author} {\bibfnamefont {Y.}~\bibnamefont {Lee}}, \bibinfo
  {author} {\bibfnamefont {Y.~W.}\ \bibnamefont {Choi}}, \bibinfo {author}
  {\bibfnamefont {T.~S.}\ \bibnamefont {Jung}}, \bibinfo {author}
  {\bibfnamefont {S.}~\bibnamefont {Son}}, \bibinfo {author} {\bibfnamefont
  {J.}~\bibnamefont {Kim}}, \bibinfo {author} {\bibfnamefont {H.~J.}\
  \bibnamefont {Choi}}, \bibinfo {author} {\bibfnamefont {J.-G.}\ \bibnamefont
  {Park}},\ and\ \bibinfo {author} {\bibfnamefont {J.~H.}\ \bibnamefont
  {Kim}},\ }\bibfield  {title} {\bibinfo {title} {Terahertz spectroscopy and
  {DFT} analysis of phonon dynamics of the layered van der waals semiconductor
  nb3x8 (x = cl, i)},\ }\href {https://doi.org/10.1021/acsomega.3c01019}
  {\bibfield  {journal} {\bibinfo  {journal} {{ACS} Omega}\ }\textbf {\bibinfo
  {volume} {8}},\ \bibinfo {pages} {14190} (\bibinfo {year}
  {2023})}\BibitemShut {NoStop}%
\bibitem [{\citenamefont {van Loon}\ \emph {et~al.}(2021)\citenamefont {van
  Loon}, \citenamefont {R\"osner}, \citenamefont {Katsnelson},\ and\
  \citenamefont {Wehling}}]{vanLoon21}%
  \BibitemOpen
  \bibfield  {author} {\bibinfo {author} {\bibfnamefont {E.~G. C.~P.}\
  \bibnamefont {van Loon}}, \bibinfo {author} {\bibfnamefont {M.}~\bibnamefont
  {R\"osner}}, \bibinfo {author} {\bibfnamefont {M.~I.}\ \bibnamefont
  {Katsnelson}},\ and\ \bibinfo {author} {\bibfnamefont {T.~O.}\ \bibnamefont
  {Wehling}},\ }\bibfield  {title} {\bibinfo {title} {Random phase
  approximation for gapped systems: Role of vertex corrections and
  applicability of the constrained random phase approximation},\ }\href
  {https://doi.org/10.1103/PhysRevB.104.045134} {\bibfield  {journal} {\bibinfo
   {journal} {Phys. Rev. B}\ }\textbf {\bibinfo {volume} {104}},\ \bibinfo
  {pages} {045134} (\bibinfo {year} {2021})}\BibitemShut {NoStop}%
\bibitem [{\citenamefont {Casula}\ \emph
  {et~al.}(2012{\natexlab{a}})\citenamefont {Casula}, \citenamefont {Rubtsov},\
  and\ \citenamefont {Biermann}}]{Casula12}%
  \BibitemOpen
  \bibfield  {author} {\bibinfo {author} {\bibfnamefont {M.}~\bibnamefont
  {Casula}}, \bibinfo {author} {\bibfnamefont {A.}~\bibnamefont {Rubtsov}},\
  and\ \bibinfo {author} {\bibfnamefont {S.}~\bibnamefont {Biermann}},\
  }\bibfield  {title} {\bibinfo {title} {Dynamical screening effects in
  correlated materials: Plasmon satellites and spectral weight transfers from a
  green's function ansatz to extended dynamical mean field theory},\ }\href
  {https://doi.org/10.1103/PhysRevB.85.035115} {\bibfield  {journal} {\bibinfo
  {journal} {Phys. Rev. B}\ }\textbf {\bibinfo {volume} {85}},\ \bibinfo
  {pages} {035115} (\bibinfo {year} {2012}{\natexlab{a}})}\BibitemShut
  {NoStop}%
\bibitem [{\citenamefont {Wietek}\ \emph {et~al.}(2021)\citenamefont {Wietek},
  \citenamefont {Rossi}, \citenamefont {\ifmmode~\check{S}\else
  \v{S}\fi{}imkovic}, \citenamefont {Klett}, \citenamefont {Hansmann},
  \citenamefont {Ferrero}, \citenamefont {Stoudenmire}, \citenamefont
  {Sch\"afer},\ and\ \citenamefont {Georges}}]{Wietek21}%
  \BibitemOpen
  \bibfield  {author} {\bibinfo {author} {\bibfnamefont {A.}~\bibnamefont
  {Wietek}}, \bibinfo {author} {\bibfnamefont {R.}~\bibnamefont {Rossi}},
  \bibinfo {author} {\bibfnamefont {F.}~\bibnamefont {\ifmmode~\check{S}\else
  \v{S}\fi{}imkovic}}, \bibinfo {author} {\bibfnamefont {M.}~\bibnamefont
  {Klett}}, \bibinfo {author} {\bibfnamefont {P.}~\bibnamefont {Hansmann}},
  \bibinfo {author} {\bibfnamefont {M.}~\bibnamefont {Ferrero}}, \bibinfo
  {author} {\bibfnamefont {E.~M.}\ \bibnamefont {Stoudenmire}}, \bibinfo
  {author} {\bibfnamefont {T.}~\bibnamefont {Sch\"afer}},\ and\ \bibinfo
  {author} {\bibfnamefont {A.}~\bibnamefont {Georges}},\ }\bibfield  {title}
  {\bibinfo {title} {Mott insulating states with competing orders in the
  triangular lattice hubbard model},\ }\href
  {https://doi.org/10.1103/PhysRevX.11.041013} {\bibfield  {journal} {\bibinfo
  {journal} {Phys. Rev. X}\ }\textbf {\bibinfo {volume} {11}},\ \bibinfo
  {pages} {041013} (\bibinfo {year} {2021})}\BibitemShut {NoStop}%
\bibitem [{\citenamefont {Lichtenstein}\ and\ \citenamefont
  {Katsnelson}(1998)}]{Lichtenstein98}%
  \BibitemOpen
  \bibfield  {author} {\bibinfo {author} {\bibfnamefont {A.~I.}\ \bibnamefont
  {Lichtenstein}}\ and\ \bibinfo {author} {\bibfnamefont {M.~I.}\ \bibnamefont
  {Katsnelson}},\ }\bibfield  {title} {\bibinfo {title} {Ab initio calculations
  of quasiparticle band structure in correlated systems: {LDA++} approach},\
  }\href {https://doi.org/10.1103/PhysRevB.57.6884} {\bibfield  {journal}
  {\bibinfo  {journal} {Phys. Rev. B}\ }\textbf {\bibinfo {volume} {57}},\
  \bibinfo {pages} {6884} (\bibinfo {year} {1998})}\BibitemShut {NoStop}%
\bibitem [{\citenamefont {Locht}\ \emph
  {et~al.}(2016{\natexlab{a}})\citenamefont {Locht}, \citenamefont {Kvashnin},
  \citenamefont {Rodrigues}, \citenamefont {Pereiro}, \citenamefont {Bergman},
  \citenamefont {Bergqvist}, \citenamefont {Lichtenstein}, \citenamefont
  {Katsnelson}, \citenamefont {Delin}, \citenamefont {Klautau}, \citenamefont
  {Johansson}, \citenamefont {Di~Marco},\ and\ \citenamefont
  {Eriksson}}]{Locht2016}%
  \BibitemOpen
  \bibfield  {author} {\bibinfo {author} {\bibfnamefont {I.~L.~M.}\
  \bibnamefont {Locht}}, \bibinfo {author} {\bibfnamefont {Y.~O.}\ \bibnamefont
  {Kvashnin}}, \bibinfo {author} {\bibfnamefont {D.~C.~M.}\ \bibnamefont
  {Rodrigues}}, \bibinfo {author} {\bibfnamefont {M.}~\bibnamefont {Pereiro}},
  \bibinfo {author} {\bibfnamefont {A.}~\bibnamefont {Bergman}}, \bibinfo
  {author} {\bibfnamefont {L.}~\bibnamefont {Bergqvist}}, \bibinfo {author}
  {\bibfnamefont {A.~I.}\ \bibnamefont {Lichtenstein}}, \bibinfo {author}
  {\bibfnamefont {M.~I.}\ \bibnamefont {Katsnelson}}, \bibinfo {author}
  {\bibfnamefont {A.}~\bibnamefont {Delin}}, \bibinfo {author} {\bibfnamefont
  {A.~B.}\ \bibnamefont {Klautau}}, \bibinfo {author} {\bibfnamefont
  {B.}~\bibnamefont {Johansson}}, \bibinfo {author} {\bibfnamefont
  {I.}~\bibnamefont {Di~Marco}},\ and\ \bibinfo {author} {\bibfnamefont
  {O.}~\bibnamefont {Eriksson}},\ }\bibfield  {title} {\bibinfo {title}
  {Standard model of the rare earths analyzed from the {Hubbard I}
  approximation},\ }\href {https://doi.org/10.1103/PhysRevB.94.085137}
  {\bibfield  {journal} {\bibinfo  {journal} {Phys. Rev. B}\ }\textbf {\bibinfo
  {volume} {94}},\ \bibinfo {pages} {085137} (\bibinfo {year}
  {2016}{\natexlab{a}})}\BibitemShut {NoStop}%
\bibitem [{\citenamefont {Pasco}\ \emph {et~al.}(2019)\citenamefont {Pasco},
  \citenamefont {El~Baggari}, \citenamefont {Bianco}, \citenamefont
  {Kourkoutis},\ and\ \citenamefont {{McQueen}}}]{pasco_tunable_2019}%
  \BibitemOpen
  \bibfield  {author} {\bibinfo {author} {\bibfnamefont {C.~M.}\ \bibnamefont
  {Pasco}}, \bibinfo {author} {\bibfnamefont {I.}~\bibnamefont {El~Baggari}},
  \bibinfo {author} {\bibfnamefont {E.}~\bibnamefont {Bianco}}, \bibinfo
  {author} {\bibfnamefont {L.~F.}\ \bibnamefont {Kourkoutis}},\ and\ \bibinfo
  {author} {\bibfnamefont {T.~M.}\ \bibnamefont {{McQueen}}},\ }\bibfield
  {title} {\bibinfo {title} {Tunable magnetic transition to a singlet ground
  state in a 2d van der waals layered trimerized kagomé magnet},\ }\href
  {https://doi.org/10.1021/acsnano.9b04392} {\bibfield  {journal} {\bibinfo
  {journal} {{ACS} Nano}\ }\textbf {\bibinfo {volume} {13}},\ \bibinfo {pages}
  {9457} (\bibinfo {year} {2019})}\BibitemShut {NoStop}%
\bibitem [{\citenamefont {Irkhin}\ \emph {et~al.}(1999)\citenamefont {Irkhin},
  \citenamefont {Katanin},\ and\ \citenamefont {Katsnelson}}]{irkhin99}%
  \BibitemOpen
  \bibfield  {author} {\bibinfo {author} {\bibfnamefont {V.~Y.}\ \bibnamefont
  {Irkhin}}, \bibinfo {author} {\bibfnamefont {A.~A.}\ \bibnamefont
  {Katanin}},\ and\ \bibinfo {author} {\bibfnamefont {M.~I.}\ \bibnamefont
  {Katsnelson}},\ }\bibfield  {title} {\bibinfo {title} {Self-consistent
  spin-wave theory of layered {Heisenberg} magnets},\ }\href
  {https://doi.org/10.1103/PhysRevB.60.1082} {\bibfield  {journal} {\bibinfo
  {journal} {Phys. Rev. B}\ }\textbf {\bibinfo {volume} {60}},\ \bibinfo
  {pages} {1082} (\bibinfo {year} {1999})}\BibitemShut {NoStop}%
\bibitem [{\citenamefont {R\"osner}\ \emph {et~al.}(2015)\citenamefont
  {R\"osner}, \citenamefont {\ifmmode \mbox{\c{S}}\else \c{S}\fi{}a\ifmmode
  \mbox{\c{s}}\else \c{s}\fi{}\ifmmode \imath \else \i
  \fi{}o\ifmmode~\breve{g}\else \u{g}\fi{}lu}, \citenamefont {Friedrich},
  \citenamefont {Bl\"ugel},\ and\ \citenamefont {Wehling}}]{roesner15}%
  \BibitemOpen
  \bibfield  {author} {\bibinfo {author} {\bibfnamefont {M.}~\bibnamefont
  {R\"osner}}, \bibinfo {author} {\bibfnamefont {E.}~\bibnamefont {\ifmmode
  \mbox{\c{S}}\else \c{S}\fi{}a\ifmmode \mbox{\c{s}}\else \c{s}\fi{}\ifmmode
  \imath \else \i \fi{}o\ifmmode~\breve{g}\else \u{g}\fi{}lu}}, \bibinfo
  {author} {\bibfnamefont {C.}~\bibnamefont {Friedrich}}, \bibinfo {author}
  {\bibfnamefont {S.}~\bibnamefont {Bl\"ugel}},\ and\ \bibinfo {author}
  {\bibfnamefont {T.~O.}\ \bibnamefont {Wehling}},\ }\bibfield  {title}
  {\bibinfo {title} {Wannier function approach to realistic coulomb
  interactions in layered materials and heterostructures},\ }\href
  {https://doi.org/10.1103/PhysRevB.92.085102} {\bibfield  {journal} {\bibinfo
  {journal} {Phys. Rev. B}\ }\textbf {\bibinfo {volume} {92}},\ \bibinfo
  {pages} {085102} (\bibinfo {year} {2015})}\BibitemShut {NoStop}%
\bibitem [{\citenamefont {van Loon}\ \emph {et~al.}(2022)\citenamefont {van
  Loon}, \citenamefont {Schüler}, \citenamefont {Springer}, \citenamefont
  {Sangiovanni}, \citenamefont {Tomczak},\ and\ \citenamefont
  {Wehling}}]{vanloon2022coulomb}%
  \BibitemOpen
  \bibfield  {author} {\bibinfo {author} {\bibfnamefont {E.~G. C.~P.}\
  \bibnamefont {van Loon}}, \bibinfo {author} {\bibfnamefont {M.}~\bibnamefont
  {Schüler}}, \bibinfo {author} {\bibfnamefont {D.}~\bibnamefont {Springer}},
  \bibinfo {author} {\bibfnamefont {G.}~\bibnamefont {Sangiovanni}}, \bibinfo
  {author} {\bibfnamefont {J.~M.}\ \bibnamefont {Tomczak}},\ and\ \bibinfo
  {author} {\bibfnamefont {T.~O.}\ \bibnamefont {Wehling}},\ }\href@noop {}
  {\bibinfo {title} {Coulomb engineering of two-dimensional mott materials}}
  (\bibinfo {year} {2022}),\ \Eprint {https://arxiv.org/abs/2001.01735}
  {arXiv:2001.01735 [cond-mat.str-el]} \BibitemShut {NoStop}%
\bibitem [{\citenamefont {Koseki}\ \emph {et~al.}(2019)\citenamefont {Koseki},
  \citenamefont {Matsunaga}, \citenamefont {Asada}, \citenamefont {Schmidt},\
  and\ \citenamefont {Gordon}}]{koseki_spinorbit_2019}%
  \BibitemOpen
  \bibfield  {author} {\bibinfo {author} {\bibfnamefont {S.}~\bibnamefont
  {Koseki}}, \bibinfo {author} {\bibfnamefont {N.}~\bibnamefont {Matsunaga}},
  \bibinfo {author} {\bibfnamefont {T.}~\bibnamefont {Asada}}, \bibinfo
  {author} {\bibfnamefont {M.~W.}\ \bibnamefont {Schmidt}},\ and\ \bibinfo
  {author} {\bibfnamefont {M.~S.}\ \bibnamefont {Gordon}},\ }\bibfield  {title}
  {\bibinfo {title} {Spin–orbit coupling constants in atoms and ions of
  transition elements: Comparison of effective core potentials, model core
  potentials, and all-electron methods},\ }\href
  {https://doi.org/10.1021/acs.jpca.8b09218} {\bibfield  {journal} {\bibinfo
  {journal} {The Journal of Physical Chemistry A}\ }\textbf {\bibinfo {volume}
  {123}},\ \bibinfo {pages} {2325} (\bibinfo {year} {2019})}\BibitemShut
  {NoStop}%
\bibitem [{\citenamefont {Locht}\ \emph
  {et~al.}(2016{\natexlab{b}})\citenamefont {Locht}, \citenamefont {Kvashnin},
  \citenamefont {Rodrigues}, \citenamefont {Pereiro}, \citenamefont {Bergman},
  \citenamefont {Bergqvist}, \citenamefont {Lichtenstein}, \citenamefont
  {Katsnelson}, \citenamefont {Delin}, \citenamefont {Klautau}, \citenamefont
  {Johansson}, \citenamefont {Di~Marco},\ and\ \citenamefont
  {Eriksson}}]{locht_standard_2016}%
  \BibitemOpen
  \bibfield  {author} {\bibinfo {author} {\bibfnamefont {I.~L.~M.}\
  \bibnamefont {Locht}}, \bibinfo {author} {\bibfnamefont {Y.~O.}\ \bibnamefont
  {Kvashnin}}, \bibinfo {author} {\bibfnamefont {D.~C.~M.}\ \bibnamefont
  {Rodrigues}}, \bibinfo {author} {\bibfnamefont {M.}~\bibnamefont {Pereiro}},
  \bibinfo {author} {\bibfnamefont {A.}~\bibnamefont {Bergman}}, \bibinfo
  {author} {\bibfnamefont {L.}~\bibnamefont {Bergqvist}}, \bibinfo {author}
  {\bibfnamefont {A.~I.}\ \bibnamefont {Lichtenstein}}, \bibinfo {author}
  {\bibfnamefont {M.~I.}\ \bibnamefont {Katsnelson}}, \bibinfo {author}
  {\bibfnamefont {A.}~\bibnamefont {Delin}}, \bibinfo {author} {\bibfnamefont
  {A.~B.}\ \bibnamefont {Klautau}}, \bibinfo {author} {\bibfnamefont
  {B.}~\bibnamefont {Johansson}}, \bibinfo {author} {\bibfnamefont
  {I.}~\bibnamefont {Di~Marco}},\ and\ \bibinfo {author} {\bibfnamefont
  {O.}~\bibnamefont {Eriksson}},\ }\bibfield  {title} {\bibinfo {title}
  {Standard model of the rare earths analyzed from the hubbard i
  approximation},\ }\href {https://doi.org/10.1103/PhysRevB.94.085137}
  {\bibfield  {journal} {\bibinfo  {journal} {Phys. Rev. B}\ }\textbf {\bibinfo
  {volume} {94}},\ \bibinfo {pages} {085137} (\bibinfo {year}
  {2016}{\natexlab{b}})}\BibitemShut {NoStop}%
\bibitem [{\citenamefont {Leb\`egue}\ \emph {et~al.}(2005)\citenamefont
  {Leb\`egue}, \citenamefont {Santi}, \citenamefont {Svane}, \citenamefont
  {Bengone}, \citenamefont {Katsnelson}, \citenamefont {Lichtenstein},\ and\
  \citenamefont {Eriksson}}]{Svane2005}%
  \BibitemOpen
  \bibfield  {author} {\bibinfo {author} {\bibfnamefont {S.}~\bibnamefont
  {Leb\`egue}}, \bibinfo {author} {\bibfnamefont {G.}~\bibnamefont {Santi}},
  \bibinfo {author} {\bibfnamefont {A.}~\bibnamefont {Svane}}, \bibinfo
  {author} {\bibfnamefont {O.}~\bibnamefont {Bengone}}, \bibinfo {author}
  {\bibfnamefont {M.~I.}\ \bibnamefont {Katsnelson}}, \bibinfo {author}
  {\bibfnamefont {A.~I.}\ \bibnamefont {Lichtenstein}},\ and\ \bibinfo {author}
  {\bibfnamefont {O.}~\bibnamefont {Eriksson}},\ }\bibfield  {title} {\bibinfo
  {title} {Electronic structure and spectroscopic properties of thulium
  monochalcogenides},\ }\href {https://doi.org/10.1103/PhysRevB.72.245102}
  {\bibfield  {journal} {\bibinfo  {journal} {Phys. Rev. B}\ }\textbf {\bibinfo
  {volume} {72}},\ \bibinfo {pages} {245102} (\bibinfo {year}
  {2005})}\BibitemShut {NoStop}%
\bibitem [{\citenamefont {Westerhout}\ and\ \citenamefont
  {Katsnelson}(2022)}]{Tom22}%
  \BibitemOpen
  \bibfield  {author} {\bibinfo {author} {\bibfnamefont {T.}~\bibnamefont
  {Westerhout}}\ and\ \bibinfo {author} {\bibfnamefont {M.~I.}\ \bibnamefont
  {Katsnelson}},\ }\bibfield  {title} {\bibinfo {title} {Role of correlated
  hopping in the many-body physics of flat-band systems: Nagaoka
  ferromagnetism},\ }\href {https://doi.org/10.1103/PhysRevB.106.L041104}
  {\bibfield  {journal} {\bibinfo  {journal} {Phys. Rev. B}\ }\textbf {\bibinfo
  {volume} {106}},\ \bibinfo {pages} {L041104} (\bibinfo {year}
  {2022})}\BibitemShut {NoStop}%
\bibitem [{\citenamefont {Peters}\ \emph {et~al.}(2014)\citenamefont {Peters},
  \citenamefont {Di~Marco}, \citenamefont {Thunstr\"om}, \citenamefont
  {Katsnelson}, \citenamefont {Kirilyuk},\ and\ \citenamefont
  {Eriksson}}]{Peters2014}%
  \BibitemOpen
  \bibfield  {author} {\bibinfo {author} {\bibfnamefont {L.}~\bibnamefont
  {Peters}}, \bibinfo {author} {\bibfnamefont {I.}~\bibnamefont {Di~Marco}},
  \bibinfo {author} {\bibfnamefont {P.}~\bibnamefont {Thunstr\"om}}, \bibinfo
  {author} {\bibfnamefont {M.~I.}\ \bibnamefont {Katsnelson}}, \bibinfo
  {author} {\bibfnamefont {A.}~\bibnamefont {Kirilyuk}},\ and\ \bibinfo
  {author} {\bibfnamefont {O.}~\bibnamefont {Eriksson}},\ }\bibfield  {title}
  {\bibinfo {title} {Treatment of $4f$ states of the rare earths: The case
  study of {TbN}},\ }\href {https://doi.org/10.1103/PhysRevB.89.205109}
  {\bibfield  {journal} {\bibinfo  {journal} {Phys. Rev. B}\ }\textbf {\bibinfo
  {volume} {89}},\ \bibinfo {pages} {205109} (\bibinfo {year}
  {2014})}\BibitemShut {NoStop}%
\bibitem [{\citenamefont {Xu}\ \emph {et~al.}(2021)\citenamefont {Xu},
  \citenamefont {Elcoro}, \citenamefont {Li}, \citenamefont {Song},
  \citenamefont {Regnault}, \citenamefont {Yang}, \citenamefont {Sun},
  \citenamefont {Parkin}, \citenamefont {Felser},\ and\ \citenamefont
  {Bernevig}}]{xu2021threedimensional}%
  \BibitemOpen
  \bibfield  {author} {\bibinfo {author} {\bibfnamefont {Y.}~\bibnamefont
  {Xu}}, \bibinfo {author} {\bibfnamefont {L.}~\bibnamefont {Elcoro}}, \bibinfo
  {author} {\bibfnamefont {G.}~\bibnamefont {Li}}, \bibinfo {author}
  {\bibfnamefont {Z.-D.}\ \bibnamefont {Song}}, \bibinfo {author}
  {\bibfnamefont {N.}~\bibnamefont {Regnault}}, \bibinfo {author}
  {\bibfnamefont {Q.}~\bibnamefont {Yang}}, \bibinfo {author} {\bibfnamefont
  {Y.}~\bibnamefont {Sun}}, \bibinfo {author} {\bibfnamefont {S.}~\bibnamefont
  {Parkin}}, \bibinfo {author} {\bibfnamefont {C.}~\bibnamefont {Felser}},\
  and\ \bibinfo {author} {\bibfnamefont {B.~A.}\ \bibnamefont {Bernevig}},\
  }\href@noop {} {\bibinfo {title} {Three-dimensional real space invariants,
  obstructed atomic insulators and a new principle for active catalytic sites}}
  (\bibinfo {year} {2021}),\ \Eprint {https://arxiv.org/abs/2111.02433}
  {arXiv:2111.02433 [cond-mat.mtrl-sci]} \BibitemShut {NoStop}%
\bibitem [{\citenamefont {Perdew}\ \emph {et~al.}(1996)\citenamefont {Perdew},
  \citenamefont {Burke},\ and\ \citenamefont
  {Ernzerhof}}]{PhysRevLett.77.3865}%
  \BibitemOpen
  \bibfield  {author} {\bibinfo {author} {\bibfnamefont {J.~P.}\ \bibnamefont
  {Perdew}}, \bibinfo {author} {\bibfnamefont {K.}~\bibnamefont {Burke}},\ and\
  \bibinfo {author} {\bibfnamefont {M.}~\bibnamefont {Ernzerhof}},\ }\bibfield
  {title} {\bibinfo {title} {{Generalized Gradient Approximation Made
  Simple}},\ }\href {https://doi.org/doi.org/10.1103/PhysRevLett.77.3865}
  {\bibfield  {journal} {\bibinfo  {journal} {Phys. Rev. Lett.}\ }\textbf
  {\bibinfo {volume} {77}},\ \bibinfo {pages} {3865} (\bibinfo {year}
  {1996})}\BibitemShut {NoStop}%
\bibitem [{\citenamefont {Bl\"ochl}(1994)}]{paw}%
  \BibitemOpen
  \bibfield  {author} {\bibinfo {author} {\bibfnamefont {P.~E.}\ \bibnamefont
  {Bl\"ochl}},\ }\bibfield  {title} {\bibinfo {title} {{Projector
  augmented-wave method}},\ }\href {https://doi.org/10.1103/PhysRevB.50.17953}
  {\bibfield  {journal} {\bibinfo  {journal} {Phys. Rev. B}\ }\textbf {\bibinfo
  {volume} {50}},\ \bibinfo {pages} {17953} (\bibinfo {year}
  {1994})}\BibitemShut {NoStop}%
\bibitem [{\citenamefont {Kresse}\ and\ \citenamefont {Joubert}(1999)}]{paw2}%
  \BibitemOpen
  \bibfield  {author} {\bibinfo {author} {\bibfnamefont {G.}~\bibnamefont
  {Kresse}}\ and\ \bibinfo {author} {\bibfnamefont {D.}~\bibnamefont
  {Joubert}},\ }\bibfield  {title} {\bibinfo {title} {From ultrasoft
  pseudopotentials to the projector augmented-wave method},\ }\href
  {https://doi.org/10.1103/PhysRevB.59.1758} {\bibfield  {journal} {\bibinfo
  {journal} {Phys. Rev. B}\ }\textbf {\bibinfo {volume} {59}},\ \bibinfo
  {pages} {1758} (\bibinfo {year} {1999})}\BibitemShut {NoStop}%
\bibitem [{\citenamefont {Kresse}\ and\ \citenamefont
  {Furthmüller}(1996)}]{KRESSE199615}%
  \BibitemOpen
  \bibfield  {author} {\bibinfo {author} {\bibfnamefont {G.}~\bibnamefont
  {Kresse}}\ and\ \bibinfo {author} {\bibfnamefont {J.}~\bibnamefont
  {Furthmüller}},\ }\bibfield  {title} {\bibinfo {title} {{Efficiency of
  ab-initio total energy calculations for metals and semiconductors using a
  plane-wave basis set}},\ }\href
  {https://doi.org/10.1016/0927-0256(96)00008-0} {\bibfield  {journal}
  {\bibinfo  {journal} {Comp. Mat. Sci.}\ }\textbf {\bibinfo {volume} {6}},\
  \bibinfo {pages} {15} (\bibinfo {year} {1996})}\BibitemShut {NoStop}%
\bibitem [{\citenamefont {Kresse}\ and\ \citenamefont
  {Furthm\"uller}(1996)}]{PhysRevB.54.11169}%
  \BibitemOpen
  \bibfield  {author} {\bibinfo {author} {\bibfnamefont {G.}~\bibnamefont
  {Kresse}}\ and\ \bibinfo {author} {\bibfnamefont {J.}~\bibnamefont
  {Furthm\"uller}},\ }\bibfield  {title} {\bibinfo {title} {{Efficient
  iterative schemes for \textit{ab initio} total-energy calculations using a
  plane-wave basis set}},\ }\href {https://doi.org/10.1103/PhysRevB.54.11169}
  {\bibfield  {journal} {\bibinfo  {journal} {Phys. Rev. B}\ }\textbf {\bibinfo
  {volume} {54}},\ \bibinfo {pages} {11169} (\bibinfo {year}
  {1996})}\BibitemShut {NoStop}%
\bibitem [{\citenamefont {Mostofi}\ \emph {et~al.}(2008)\citenamefont
  {Mostofi}, \citenamefont {Yates}, \citenamefont {Lee}, \citenamefont {Souza},
  \citenamefont {Vanderbilt},\ and\ \citenamefont {Marzari}}]{MOSTOFI2008685}%
  \BibitemOpen
  \bibfield  {author} {\bibinfo {author} {\bibfnamefont {A.~A.}\ \bibnamefont
  {Mostofi}}, \bibinfo {author} {\bibfnamefont {J.~R.}\ \bibnamefont {Yates}},
  \bibinfo {author} {\bibfnamefont {Y.-S.}\ \bibnamefont {Lee}}, \bibinfo
  {author} {\bibfnamefont {I.}~\bibnamefont {Souza}}, \bibinfo {author}
  {\bibfnamefont {D.}~\bibnamefont {Vanderbilt}},\ and\ \bibinfo {author}
  {\bibfnamefont {N.}~\bibnamefont {Marzari}},\ }\bibfield  {title} {\bibinfo
  {title} {{wannier90: A tool for obtaining maximally-localised Wannier
  functions}},\ }\href {https://doi.org/10.1016/j.cpc.2007.11.016} {\bibfield
  {journal} {\bibinfo  {journal} {Comp. Phys. Commun.}\ }\textbf {\bibinfo
  {volume} {178}},\ \bibinfo {pages} {685 } (\bibinfo {year}
  {2008})}\BibitemShut {NoStop}%
\bibitem [{\citenamefont {Kaltak}(2015)}]{KaltakcRPA}%
  \BibitemOpen
  \bibfield  {author} {\bibinfo {author} {\bibfnamefont {M.}~\bibnamefont
  {Kaltak}},\ }\href {http://othes.univie.ac.at/38099/} {\bibinfo {title}
  {Merging {GW} with {DMFT}}} (\bibinfo {year} {2015}),\ \bibinfo {note} {{PhD
  Thesis, University of Vienna, 2015, 231 pp.}}\BibitemShut {Stop}%
\bibitem [{\citenamefont {Casula}\ \emph
  {et~al.}(2012{\natexlab{b}})\citenamefont {Casula}, \citenamefont {Werner},
  \citenamefont {Vaugier}, \citenamefont {Aryasetiawan}, \citenamefont
  {Miyake}, \citenamefont {Millis},\ and\ \citenamefont
  {Biermann}}]{PhysRevLett.109.126408}%
  \BibitemOpen
  \bibfield  {author} {\bibinfo {author} {\bibfnamefont {M.}~\bibnamefont
  {Casula}}, \bibinfo {author} {\bibfnamefont {P.}~\bibnamefont {Werner}},
  \bibinfo {author} {\bibfnamefont {L.}~\bibnamefont {Vaugier}}, \bibinfo
  {author} {\bibfnamefont {F.}~\bibnamefont {Aryasetiawan}}, \bibinfo {author}
  {\bibfnamefont {T.}~\bibnamefont {Miyake}}, \bibinfo {author} {\bibfnamefont
  {A.~J.}\ \bibnamefont {Millis}},\ and\ \bibinfo {author} {\bibfnamefont
  {S.}~\bibnamefont {Biermann}},\ }\bibfield  {title} {\bibinfo {title}
  {Low-energy models for correlated materials: Bandwidth renormalization from
  coulombic screening},\ }\href
  {https://doi.org/10.1103/PhysRevLett.109.126408} {\bibfield  {journal}
  {\bibinfo  {journal} {Phys. Rev. Lett.}\ }\textbf {\bibinfo {volume} {109}},\
  \bibinfo {pages} {126408} (\bibinfo {year} {2012}{\natexlab{b}})}\BibitemShut
  {NoStop}%
\bibitem [{\citenamefont {Soriano}\ \emph {et~al.}(2021)\citenamefont
  {Soriano}, \citenamefont {Rudenko}, \citenamefont {Katsnelson},\ and\
  \citenamefont {Rösner}}]{soriano_environmental_2021}%
  \BibitemOpen
  \bibfield  {author} {\bibinfo {author} {\bibfnamefont {D.}~\bibnamefont
  {Soriano}}, \bibinfo {author} {\bibfnamefont {A.~N.}\ \bibnamefont
  {Rudenko}}, \bibinfo {author} {\bibfnamefont {M.~I.}\ \bibnamefont
  {Katsnelson}},\ and\ \bibinfo {author} {\bibfnamefont {M.}~\bibnamefont
  {Rösner}},\ }\bibfield  {title} {\bibinfo {title} {Environmental screening
  and ligand-field effects to magnetism in {CrI}3 monolayer},\ }\href
  {https://doi.org/10.1038/s41524-021-00631-4} {\bibfield  {journal} {\bibinfo
  {journal} {npj Computational Materials}\ }\textbf {\bibinfo {volume} {7}},\
  \bibinfo {pages} {1} (\bibinfo {year} {2021})}\BibitemShut {NoStop}%
\bibitem [{\citenamefont {Parcollet}\ \emph {et~al.}(2015)\citenamefont
  {Parcollet}, \citenamefont {Ferrero}, \citenamefont {Ayral}, \citenamefont
  {Hafermann}, \citenamefont {Krivenko}, \citenamefont {Messio},\ and\
  \citenamefont {Seth}}]{triqs}%
  \BibitemOpen
  \bibfield  {author} {\bibinfo {author} {\bibfnamefont {O.}~\bibnamefont
  {Parcollet}}, \bibinfo {author} {\bibfnamefont {M.}~\bibnamefont {Ferrero}},
  \bibinfo {author} {\bibfnamefont {T.}~\bibnamefont {Ayral}}, \bibinfo
  {author} {\bibfnamefont {H.}~\bibnamefont {Hafermann}}, \bibinfo {author}
  {\bibfnamefont {I.}~\bibnamefont {Krivenko}}, \bibinfo {author}
  {\bibfnamefont {L.}~\bibnamefont {Messio}},\ and\ \bibinfo {author}
  {\bibfnamefont {P.}~\bibnamefont {Seth}},\ }\bibfield  {title} {\bibinfo
  {title} {Triqs: A toolbox for research on interacting quantum systems},\
  }\href {https://doi.org/http://dx.doi.org/10.1016/j.cpc.2015.04.023}
  {\bibfield  {journal} {\bibinfo  {journal} {Computer Physics Communications}\
  }\textbf {\bibinfo {volume} {196}},\ \bibinfo {pages} {398 } (\bibinfo {year}
  {2015})}\BibitemShut {NoStop}%
\bibitem [{sch(2022)}]{schueler_hubbardi_2022}%
  \BibitemOpen
  \href {https://github.com/TRIQS/hubbardI} {\bibinfo {title} {{hubbardI} - a
  hubbard-i solver based on triqs atom\_diag}} (\bibinfo {year}
  {2022})\BibitemShut {NoStop}%
\end{thebibliography}%

\end{document}